\documentclass[prl,twocolumn,showpacs,superscriptaddress,nofootinbib]{revtex4-2}%
\pdfoutput=1
\usepackage{graphicx}% Include figure files
\usepackage{epstopdf}
\usepackage{bm}% bold math
\usepackage{amssymb}
\usepackage{color}
\usepackage{amsmath}
\usepackage{amstext}
\usepackage{latexsym}
\usepackage[usenames,dvipsnames]{xcolor}
\usepackage[colorlinks=true,citecolor=MidnightBlue,linkcolor=RubineRed,urlcolor=MidnightBlue]{hyperref}

\def\x{{\rm\bf x}}

\newcommand{\beq}{\begin{equation}}
\newcommand{\eeq}{\end{equation}}
\newcommand{\beqa}{\begin{eqnarray}}
\newcommand{\eeqa}{\end{eqnarray}}

\usepackage{tikz,xcolor,hyperref}
\definecolor{lime}{HTML}{A6CE39}
\DeclareRobustCommand{\orcidicon}{
\begin{tikzpicture}
\draw[lime, fill=lime] (0,0)
circle[radius=0.16]
node[white]{{\fontfamily{qag}\selectfont \tiny \.{I}D}}; 
\end{tikzpicture}
\hspace{-2mm}
}
\foreach \x in {A, ..., Z}{%
\expandafter\xdef\csname orcid\x\endcsname{\noexpand\href{https://orcid.org/\csname orcidauthor\x\endcsname}{\noexpand\orcidicon}}
}

\begin{document}
\title{Universal defect density scaling in an oscillating dynamic phase transition}

\author{Wei-Can Yang\hspace{-1.5mm}\orcidA{}%\href{https://orcid.org/0000-0003-1520-6960}{\includegraphics[scale=0.05]{orcidid.pdf}}
}
\affiliation{Department of Physics, Osaka Metropolitan University, 3-3-138 Sugimoto, 558-8585 Osaka, Japan}
\author{Makoto Tsubota}\email{tsubota@omu.ac.jp}
\affiliation{Department of Physics, Osaka Metropolitan University, 3-3-138 Sugimoto, 558-8585 Osaka, Japan}
\affiliation{Nambu Yoichiro Institute of Theoretical and Experimental Physics (NITEP), Osaka Metropolitan University, 3-3-138 Sugimoto, Sumiyoshi-ku, Osaka 558-8585, Japan}
\author{Adolfo del Campo\hspace{-1.5mm}\orcidB{}%\href{https://orcid.org/0000-0003-2219-2851}{\includegraphics[scale=0.05]{orcidid.pdf}}
}\email{adolfo.delcampo@uni.lu}
\affiliation{Department  of  Physics  and  Materials  Science,  University  of  Luxembourg,  L-1511  Luxembourg, Luxembourg}
\affiliation{Donostia International Physics Center,  E-20018 San Sebasti\'an, Spain}
\author{Hua-Bi Zeng\hspace{-1.5mm}\orcidC{}%\href{https://orcid.org/0000-0001-7409-0537}{\includegraphics[scale=0.05]{orcidid.pdf}}
}\email{hbzeng@yzu.edu.cn}
\affiliation{Center for Theoretical Physics , Hainan University, Haikou 570228, China}
\affiliation{Center for Gravitation and Cosmology, College of Physical Science and Technology, Yangzhou University, Yangzhou 225009, China}

\begin{abstract}
 Universal scaling laws govern the density of topological defects generated while crossing an equilibrium continuous phase transition.  The Kibble-Zurek mechanism (KZM) predicts the dependence on the quench time for slow quenches.   By contrast, for fast quenches, the defect density scales universally with the amplitude of the quench.  We show that universal scaling laws apply to dynamic phase transitions driven by an oscillating external field.  The difference in the energy response of the system to a periodic potential field leads to energy absorption, spontaneous breaking of symmetry, and its restoration. We verify the associated universal scaling laws, providing evidence that the critical behavior of non-equilibrium phase transitions can be described by time-average critical exponents combined with the KZM. Our results demonstrate that the universality of critical dynamics extends beyond equilibrium criticality,  facilitating the understanding of complex non-equilibrium systems. 
 \end{abstract}
 
\maketitle

\section{Introduction}
The Kibble-Zurek Mechanism (KZM) provides a universal paradigm for the description of critical phenomena exhibited by complex systems during spontaneous symmetry breaking \cite{kibble1976topology,zurek1985cosmological,delcampo14}.  This mechanism describes the formation of long-lived topological defects when a system undergoes a continuous or quantum phase transition across a critical point at a finite rate.  While in its genesis it aimed at describing structure formation in a cosmology \cite{kibble1976topology}, the KZM has been generalized to include condensed matter system and various other scenarios \cite{zurek1985cosmological,chuang1991cosmology,mielenz2013trapping,lamporesi2013spontaneous,polkovnikov2011colloquium,weiler2008spontaneous,dziarmaga2008winding,kolodrubetz2012nonequilibrium,delcampo14}.  Its broad applicability has made it an invaluable tool for studying phase transitions and understanding the dynamics of critical phenomena in classical and quantum systems.

Consider a system with a continuous phase transition at the critical point $\lambda_c$.  In terms of the control parameter $\lambda$, we define a reduced distance parameter
   $ \epsilon=1-\lambda/\lambda_c$ that governs then the divergence  of the equilibrium correlation length $\xi$ and the equilibrium relaxation time $\tau$ 
\begin{eqnarray}
    \xi = \frac{\xi_0}{|\epsilon|^\nu},\qquad          
    \tau = \frac{\tau_0}{|\epsilon|^{z\nu}},
\end{eqnarray}
where $\nu$ and $z$  are the correlation-length and dynamic critical exponent, respectively.
Both $\nu$ and $z$ depend on the dimensionality and symmetry of the order parameter and are universal, i.e.,  independent of the microscopic details of the system \cite{HohenbergHalperin77}. 

When the control parameter is driven in time, near the critical point $\lambda_c$, critical slowing down prevents the system from adapting to the instantaneous equilibrium configuration for  $\lambda(t)$, effectively freezing the system.  KZM characterizes the non-equilibrium state resulting from a  linearized quench with $\lambda(t)=\lambda_c[1-\epsilon(t)]$ and $\epsilon(t)=t/\tau_Q$.  It makes use of the freeze-out time $\hat{t}$, estimated by comparing the relaxation time with the time gone by after crossing the critical point $\hat{t}=\tau(\hat{t})=\tau_0/|\epsilon(\hat{t})|^{z\nu}$, which yields 
$ \hat{t} \sim \tau_0(\tau_Q/\tau_0)^{z\nu/(1+z\nu)}$. KZM
predicts that  domains form  with an average size set by the equilibrium correlation length at the freeze-out time, 
$\hat{\xi}=\xi[\hat{\epsilon}]=\xi_0(\tau_Q/\tau_0)^{\nu/(1+z\nu)}$.
As a result, the average number of point-like topological defects in a $d$-dimensional system scales as
\begin{equation}
    n \sim \frac{1}{\hat{\xi}^{d}} \sim \tau_Q^{-\frac{d\nu}{1+z\nu}}.
\end{equation}

This universal scaling law predicts the relationship between the topological defects and the quench rate in systems across vastly different scales,  from cosmological scenarios to quantum phase transitions.  Supporting evidence has been found in many experiments \cite{delcampo14}, using liquid crystals \cite{chuang1991cosmology,bowick1994cosmological}, superfluid helium \cite{hendry1994cosmological,ruutu1996vortex,dodd1998nonappearance}, ultracold gases \cite{sadler2006spontaneous,weiler2008spontaneous,navon2015critical} and programmable quantum devices \cite{Cui16,Keesling19,Cui20,Bando20,King22}. While many of these experiments concern point-like defects, the KZM has also been verified in the formation of Ising domains \cite{Du2023}.

KZM was conceived to describe the dynamics across an equilibrium phase transition, combining equilibrium scaling relations with the driving of the control parameter. 
However, numerical simulations and experiments have recently demonstrated that KZM can be applied to some non-equilibrium phase transitions involving steady states  
\cite{Ducci99,Casado01,Casado06,Casado07,Miranda12,Miranda13}.  Such scenarios are characterized by nonconservation of energy and particle number and arise for instance in exciton-polaritons systems \cite{zamora2020kibble}, and directed percolation  \cite{reichhardt2022kibble,maegochi2022kibble}.  These results thus extend the universality of KZM  to non-equilibrium systems in the limit of slow quenches.

\begin{figure}[t]
    \centering
    \includegraphics[width=9.3cm,trim=20 0 0 0]{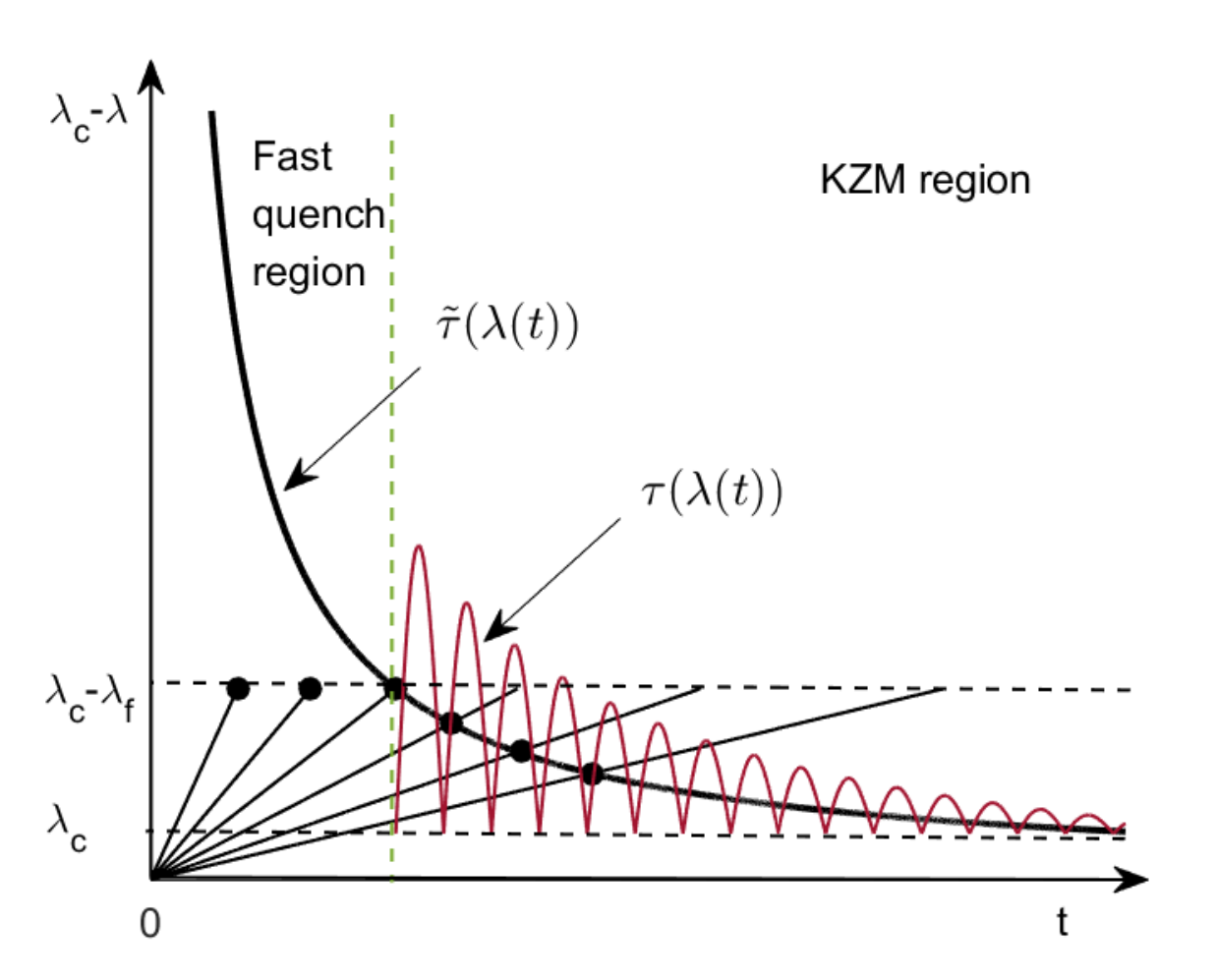}
    \caption{ Universal breakdown of KZM in a fast non-equilibrium phase transition. The effective relaxation time $\widetilde{\tau}(\lambda(t))$ should be the time-average of the real relaxation time $\tau(\lambda(t))$ (red line). The freeze-out time determined by equating the time elapsed after crossing the phase transition to average relaxation time (large black dot). In fast quench region with finite depth $\lambda_f$, the quench is saturated at a fixed equilibrium relaxation time $\widetilde{\tau}(\lambda_f)$, which we will explain later.}
    \label{figure0}
\end{figure}

Here, we consider universal scaling laws with arbitrary quench rates in another large class of non-equilibrium phase transitions: the periodic field-oscillatory dynamic phase transitions, which have been studied in a wide range of systems \cite{sides1998kinetic,tome1990dynamic,yuksel2012nonequilibrium, PhysRevResearch.5.023014,PhysRevLett.95.260404, PhysRevResearch.5.021001}.
The unique property of this kind of non-equilibrium phase transition is that the periodic field phase transition requires a continuous input of energy from the external field to maintain the phase transition state. In this phase transition, all the quantities change periodically. For example,  the relaxation time  varies according to $\tau= \frac{\tau_0}{|1- \lambda |\cos(\Omega t)|/\lambda_c|^{z\nu}}$. Although detailed balance is not satisfied, it is believed that an average critical exponent can represent the phase transition behavior \cite{zeng2018universal}. Even if the relaxation time oscillates continuously, 
%$around $\tau_0$, 
the freezing of the system can be determined by the average relaxation time $\widetilde{\tau}$,
making it possible to verify the universal scaling laws with these average critical exponents, see Fig. \ref{figure0}. An analysis of the the quasinormal modes (QNMs) can provide the real relaxation time with periodic oscillation. However, whether the critical behavior of periodically-driven phase transitions can be described by mean critical exponents is controversial. In this non-equilibrium system, we investigate whether KZM is satisfied for slow quenches and whether the number of defects under fast quenches exhibits a universal scaling behavior with the quench rate or the final value of the control parameter. We establish the validity of the nonequilibrium scaling relations with the average critical exponents.

We focus on the response of a superconductor or superfluid to an applied periodic field that is governed by the proximity of the applied field to the critical magnetic field of the superconductor or the critical velocity of the superfluid \cite{holczer1991critical,langer1967intrinsic}.
The response is bidirectional in the periodic field system.  Only when the suppression rate of the increasing field is the same as the recovery rate of the decreasing field, the system reaches an non-equilibrium steady state. 
To investigate the significant response of this order parameter to the external field \cite{herzog2009holographic,nakano2008critical,yang2023motion}, we use a holographic setting, the so-called AdS/CFT duality, for numerical experiments.  This duality arises from Maldacena's celebrated conjecture in string theory \cite{maldacena1999large} and establishes a surprising equivalence relationship between the high dimensional classical gravitational field and the low dimensional strongly coupled quantum field without gravity \cite{gubser1998gauge,witten1998anti}.  Holographic duality offers a novel first-principles perspective in condensed matter physics and finds extensive applications in the exploration of quantum many-body systems exhibiting superfluidity, superconductivity, and supersolid behavior \cite{herzog2009lectures,hartnoll2008building,baggioli2022holographic,yang2021phase}.

\section{Holographic setup and Numerical method}
Holographic duality has been extensively employed in studying KZM in equilibrium phase transitions \cite{sonner2015universal,chesler2015defect,zeng2023universal,xia2020winding}.  In particular, recent research using holographic models has uncovered a new universal scaling behavior in the fast quenching regime beyond KZM \cite{xia2021kibble}.  This behavior has been simultaneously validated in both the $\phi^4$ model and the quantum Ising chain model \cite{zeng2023universal}, further highlighting the significant importance of holographic models in providing a deeper understanding of condensed matter physics and emphasizing their relevance in explaining complex phenomena.

We employ a bottom-up holographic model, in which  $(2+1)$-dimensional superconductivity/superfluidity is described by an Abelian-Higgs model living in a $(3+1)$-dimensional asymptotically anti-de Sitter (AdS) black hole spacetime.  In this model, the AdS black hole couples with a charged scalar field and a U(1) gauge field.  Consequently, the action can be written as 
\begin{equation}
    S=\int d^4x \sqrt{-g} \big[-\frac{1}{4}F_{\mu\nu} F^{\mu\nu}-|\partial_\mu \Psi-iqA_\mu \Psi|^2-m^2|\Psi|^2 \big].
\end{equation}
Here, $F_{\mu\nu}=\partial_\mu A_\nu-\partial_\nu A_\mu$ is the Maxwell field strength with vector potential $A_{\mu}$, that is coupled minimally to the scalar involving the charge $q$.  $\Psi$ is the complex scalar field with mass  $m^2=-2$.  The background metric  in the Eddington coordinate is $ds^2=\frac{L}{z^2}[-f(z)dt^2-2dtdz+dx^2+dy^2]$, with $f(z) =1-(z/z_h)^3$ and  the Hawking temperature can be expressed as $T=3/(4\pi z_h)$. We set a square boundary with  periodic boundary conditions, where $x,y$ are the spatial coordinates, and $z$ is the extra radial dimension of the bulk.
From the static holographic superconductor \cite{hartnoll2008building}, the temperature can be controlled by the only dimensional parameter, the chemical potential $\mu$, that satisfies $T/T_c=\mu_c/\mu$. As the chemical potential grows beyond a critical value $\mu_c=4.07$, which means the temperature falls below the critical temperature, the dual scalar field at the boundary exhibits a finite expectation value, i.e., the superconductor/superfluid order parameter $\langle O \rangle$.
In this work, we fix the background temperature $T=0.77T_c$, so we have a finite order parameter in the initial state without an external field.

 Dynamic simulations can be conducted through numerical solutions of the equations of motion (EOM). These equations govern the behavior of bulk gauge and scalar fields, describing the equilibrium geometry within the bulk. They can be succinctly expressed as follows:
\begin{eqnarray}
d_\nu F^{\mu\nu} = J^{\mu} = iq(\Psi^* D^{\mu}\Psi-\Psi D^{\mu}\Psi^*), \\
(-D^2+m^2)\Psi=0.
\end{eqnarray}
By imposing holographic periodic boundary conditions, we numerically solved these equations of motion. Specifically, we use high-order Runge-Kutta methods for accuracy and efficiency. Furthermore, in the $z$-direction, we utilize the Chebyshev method, while in the $x$ and $y$ directions, the Fourier method was applied to address boundary conditions. For a $40 \times 40$ size system, we used a sufficiently accurate grid point of $300 \times 300$. 

\begin{figure}[t]
    \centering
    \includegraphics[width=9.3cm,trim=20 0 0 0]{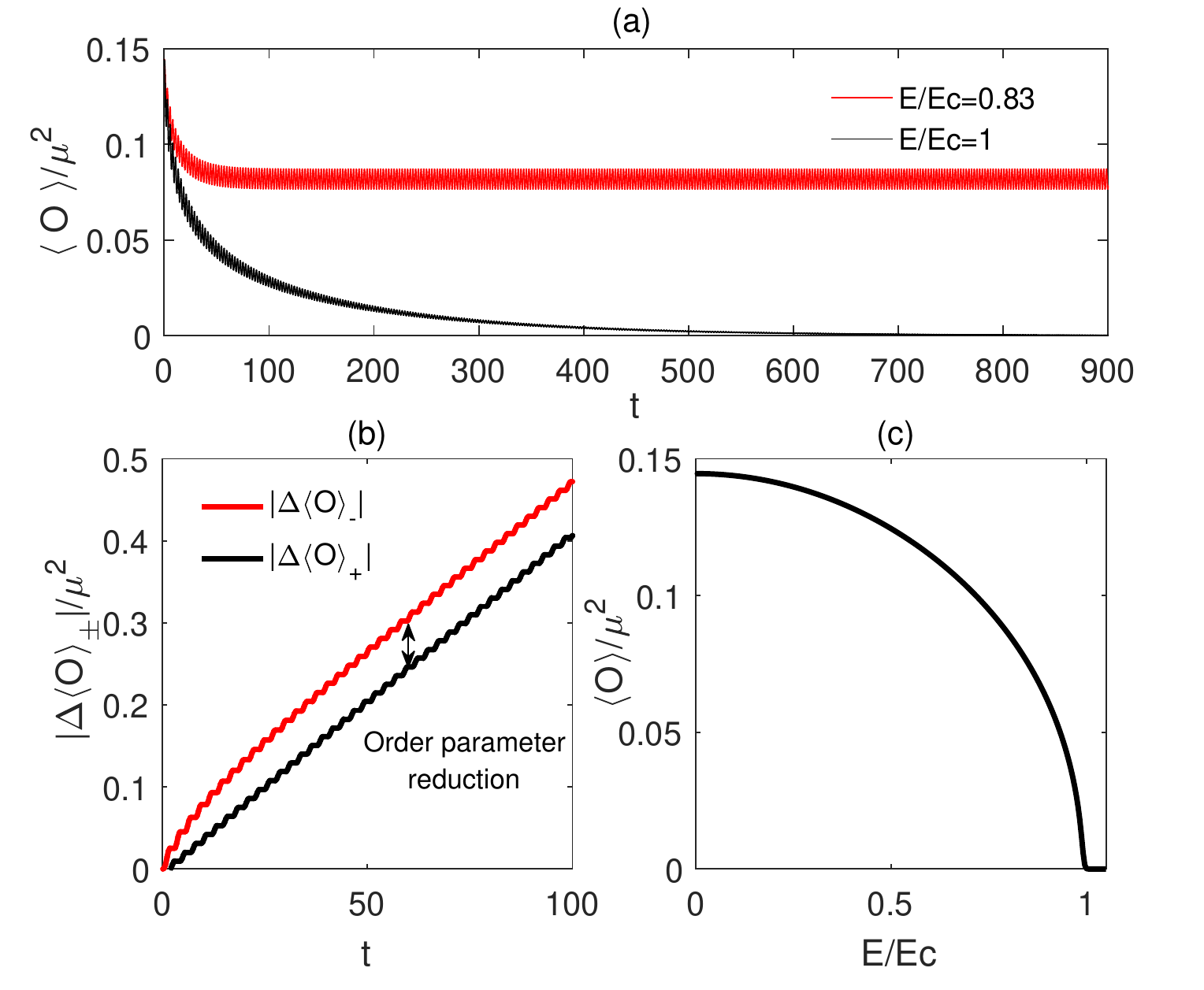}
    \caption{(a) Evolution of order parameters in an external oscillating field.  With the addition of the oscillating field, the order parameters gradually decrease and finally reach an non-equilibrium steady state.  The red line has an amplitude lower than the critical field $E/E_c=0.83$ and corresponds to an oscillatory steady state.  The black line has an amplitude equal to the critical field $E/E_c=1$ and shows the decay of the order parameter in the steady state. 
    (b) In the case of $ E/E_c=0.83$, the red line represents the rate at which the order parameters fall in response to a large electric field, while the black line represents the rate at which the small electric field order parameters recover to rise.  
    (c) Phase diagram of the final average order parameter as a function of the amplitude of the field.  When the amplitude of the field is greater than the critical value, the system is completely disordered and enters the  normal state.}
    \label{figure1}
\end{figure}

\section{Kibble-Zurek Mechanism in oscillating dynamic phase transition}
\subsection{Oscillating dynamic phase transition}

We turn on the spatial gauge field component $A_{x,y}$ to induce the external driving force into the system.  We take a sinusoidal form of the boundary conditions at $z=0$,  
\begin{equation}
    A_x(t,z=0) = \frac{E\sin(\Omega t)}{\Omega}.
\end{equation}
The oscillating field along the $x$ direction can be represented by the time derivative of the gauge field
\begin{equation}
    E_x(t) = \partial_t A_x = E\cos(\Omega t).
\end{equation}
Here, $E$ is the amplitude, and $\Omega$ is the frequency of the applied field. 
With the fixed frequency $\Omega=0.3\mu_c$, the finite amplitude $E$ causes the system to enter a non-equilibrium oscillatory state, and the average order parameter decreases relative to the stable equilibrium state.  This is an inevitable consequence of the equation of motion \cite{li2013periodically}. 

\begin{figure}[t]
    \centering
    \includegraphics[width=9.1cm,trim=30 0 0 0]{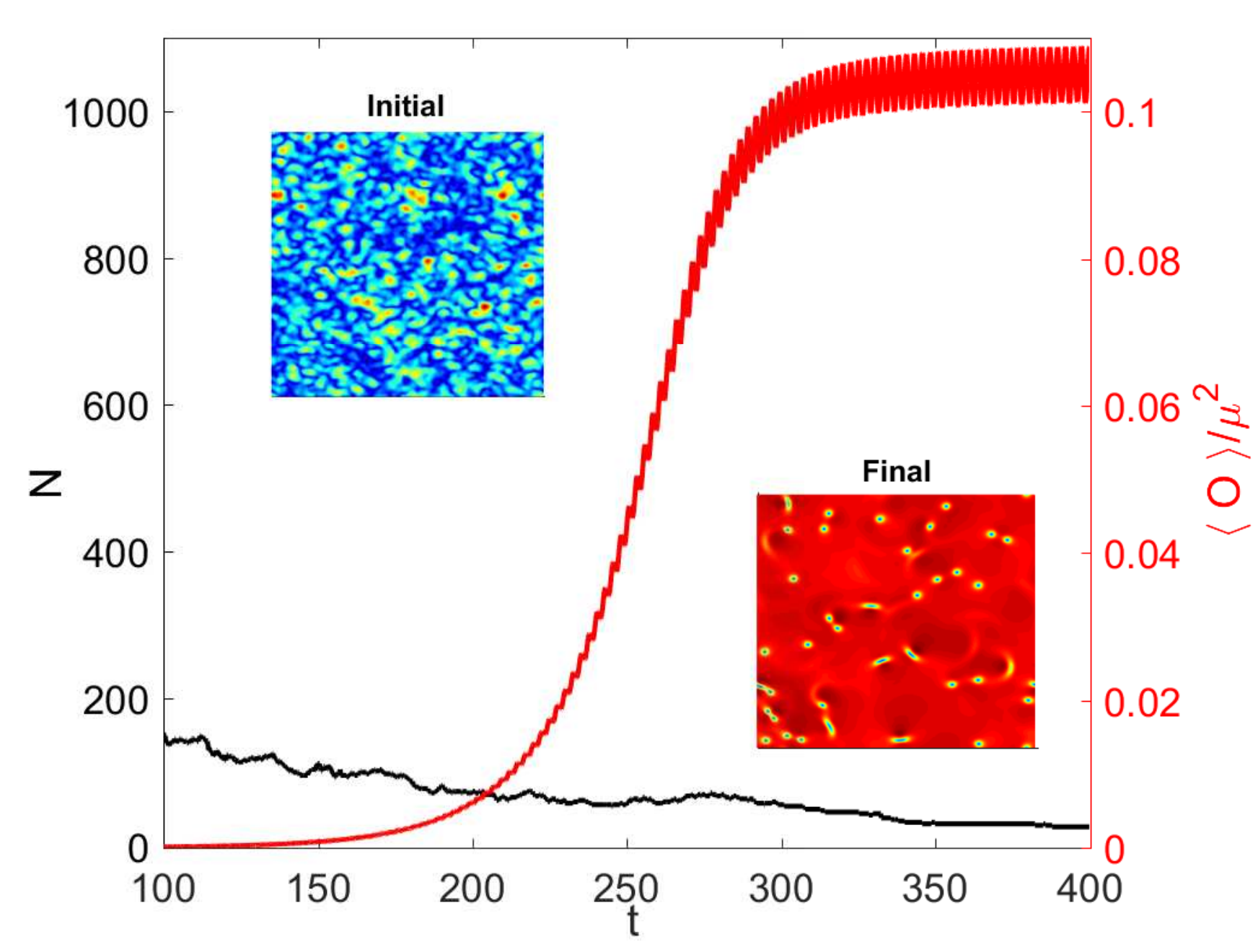}
    \caption{Dynamic evolution of the non-equilibrium system with topological defects generated after a finite time quench with an oscillating field.  The solid black line shows the number of topological defects with noise.  The red line shows the evolution of the order parameter. 
    The insets show the profile of order parameters before and after the freeze-out time, manifesting a distribution of vortices in the latter stage. }
    %in the frozen and adiabatic stages, respectively.}
    \label{figure2}
\end{figure}

As shown in the upper panel in Fig. \ref{figure1}, when an oscillating field is added to the initial statically stable solution, the order parameters begin to oscillate, and the average value decreases gradually. 
Eventually, according to the amplitude of the field, a non-equilibrium steady state with a different mean value is reached.  Our result shows that the critical value of the amplitude is $E_c = 0.705\mu_c$.
The decay of the order parameters is related to the system's response to the external field.

The bottom left panel shows separately the increasing and decreasing velocities of the order parameters, defined as 
$|\Delta\langle O \rangle _{\pm}|$, with the rates defined as $v_\pm =\partial_t |\Delta\langle O \rangle _{\pm}|$. Since the decrease of the order parameter is accompanied by energy absorption, we refer to  $v_-$ as the absorption rate and $v_+$ as the release rate.  At the beginning, when the order parameter is large enough, the system's response to the large field is significant, so the absorption rate is much larger than the release rate.  With the decrease of the order parameter, the absorption rate gradually declines and finally becomes equal to the release rate.  At this time, the total reduction of the order parameter can be obtained from the integral difference between the two rates. 
The bottom right panel shows the complete phase diagram for this fixed temperature and frequency.  The corresponding value of each field intensity is the long-time average value of the order parameter at the non-equilibrium steady state.
Therefore, although detailed balance cannot be satisfied,  the critical behavior of this non-equilibrium phase transition can be expressed in terms of the average critical exponent.

\subsection{Quenching process}

To linearly traverse the critical point and establish the universal scaling behavior in the critcial dynamics, we first prepare an initial state where the electric field is slightly larger than the critical value $E_{i}=E_c+dE_c$, and then linearly quench to an ordered phase with an electric field of $E_f<E_c$ in a finite quench time $\tau_Q$. In doing so,  $E(t)= E_c [1-\epsilon(t)]$ and $\epsilon(t)=t/\tau_Q$.
To dynamically break system symmetry, we introduce the noise  into the scalar field's evolution process within the bulk by satisfying the statistical distributions $\langle s(t, x_i) \rangle=0$ and $\langle s(t, x_i)s(t', x_j ) \rangle =h\delta(t - t')\delta(x_i - x_j )$ for every 100 time steps, with the amplitude $h = 10^3$ \cite{Sonner_2015,Zeng_2021}. 
With the linear decrease of the amplitude of the electric field $E$, after crossing the frozen stage in the KZM description set by the freeze-out time, the system responds in the opposite direction to the previous description in Fig. \ref{figure1}, resulting in non-equilibrium symmetry breaking, and the order parameters gradually increase.  As shown in Fig. \ref{figure2}, where we set $\tau_Q=120$, the order parameters grow in an oscillatory fashion.
At the same time, stable topological defects begin to appear.  The distribution of topological defects can be identified in the built-in diagram at the lower right panel.
 We calculate the number of vortices by computing the phase winding numbers $N = \frac{1}{2\pi} \oint \nabla\theta \cdot dr$, where $\theta$ is the phase of the wave function which is defined as $\langle O \rangle=\psi = |\psi| e^{i\theta}$. 
To rule out contributions dominated by noise, the number of topological defects is determined after the order parameter starts to grow,  acquiring a finite value.  For each $\tau_Q$, we fix the specific  value  $\langle O \rangle =0.1$  and count the number of topological defects when this value of the order parameter is first reached $N_{t_f}=N_{\langle O \rangle =0.1}$. The complete evolution process is shown in the Movie in the supplementary file \cite{supplementary}. 

\begin{figure}[t]
%    \centering
    \includegraphics[width=9.3cm,trim=0 0 0 0]{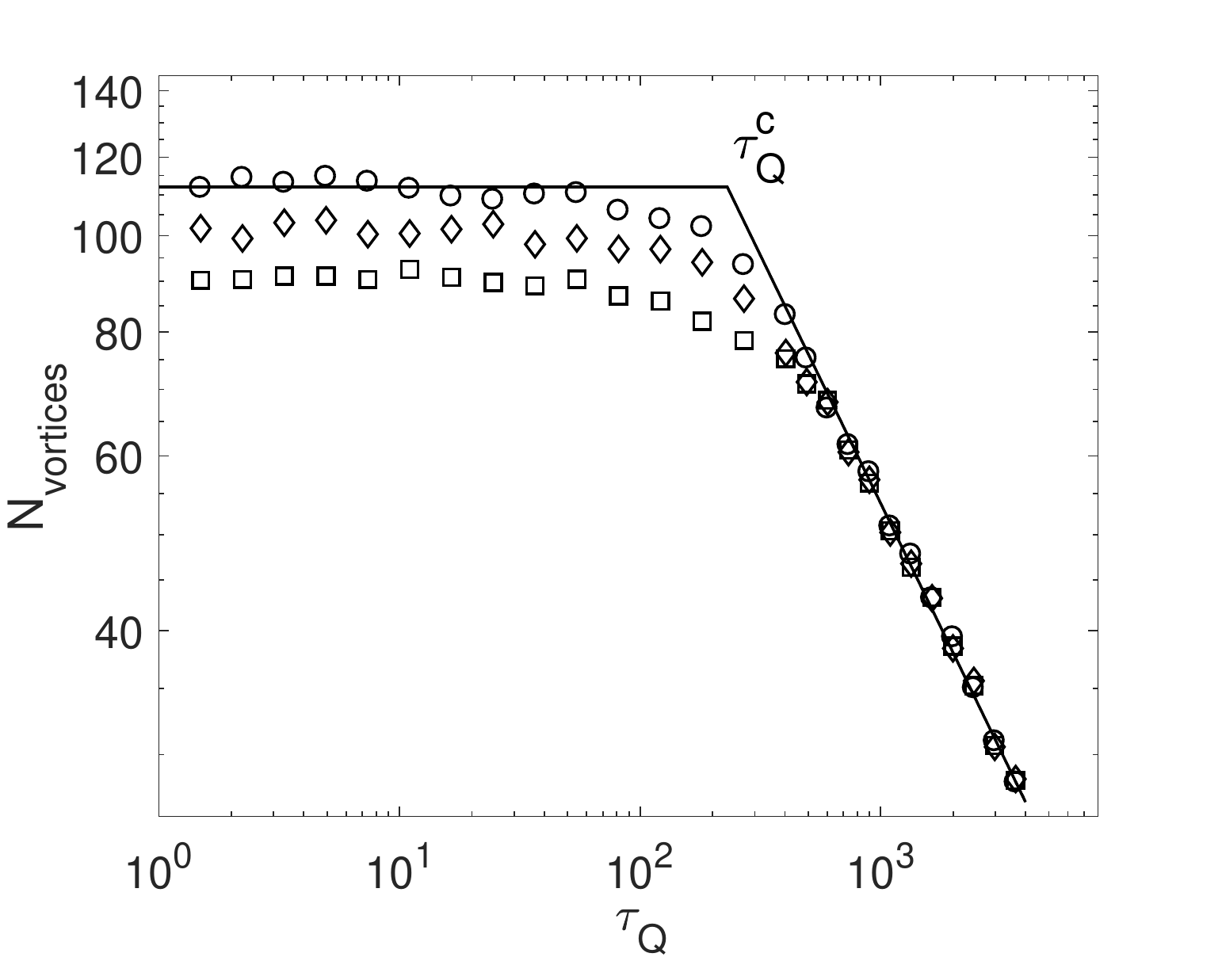}
    \caption{The number of vortices $N_{vortices}$ as a function of the quench time $\tau_Q$ in which the the electric field decays. For each $\tau_Q$ point, the number of topological defects is averaged over 100 trajectories.  For slow quenches with  $\tau_Q>200$, the vortex number is consistent with the KZM scaling $N \sim \tau_Q^{-0.4832\pm0.0182}$. 
    For fast quenches with  $\tau_Q < 200$, the KZM is no longer satisfied, and the number of vortices does not vary with the quench rate.  However,  the value at the plateau scales universally with the amplitude of the quench, as shown from top to bottom, for the values of $E_f=0E_c, 0.4E_c, 0.7Ec$, with the symbols $\bigcirc, \Diamond, \square$, respectively.}
    \label{figure3}
\end{figure}

\begin{figure*}[t]
    \centering
    \includegraphics[width=15cm,trim=0 50 0 40]{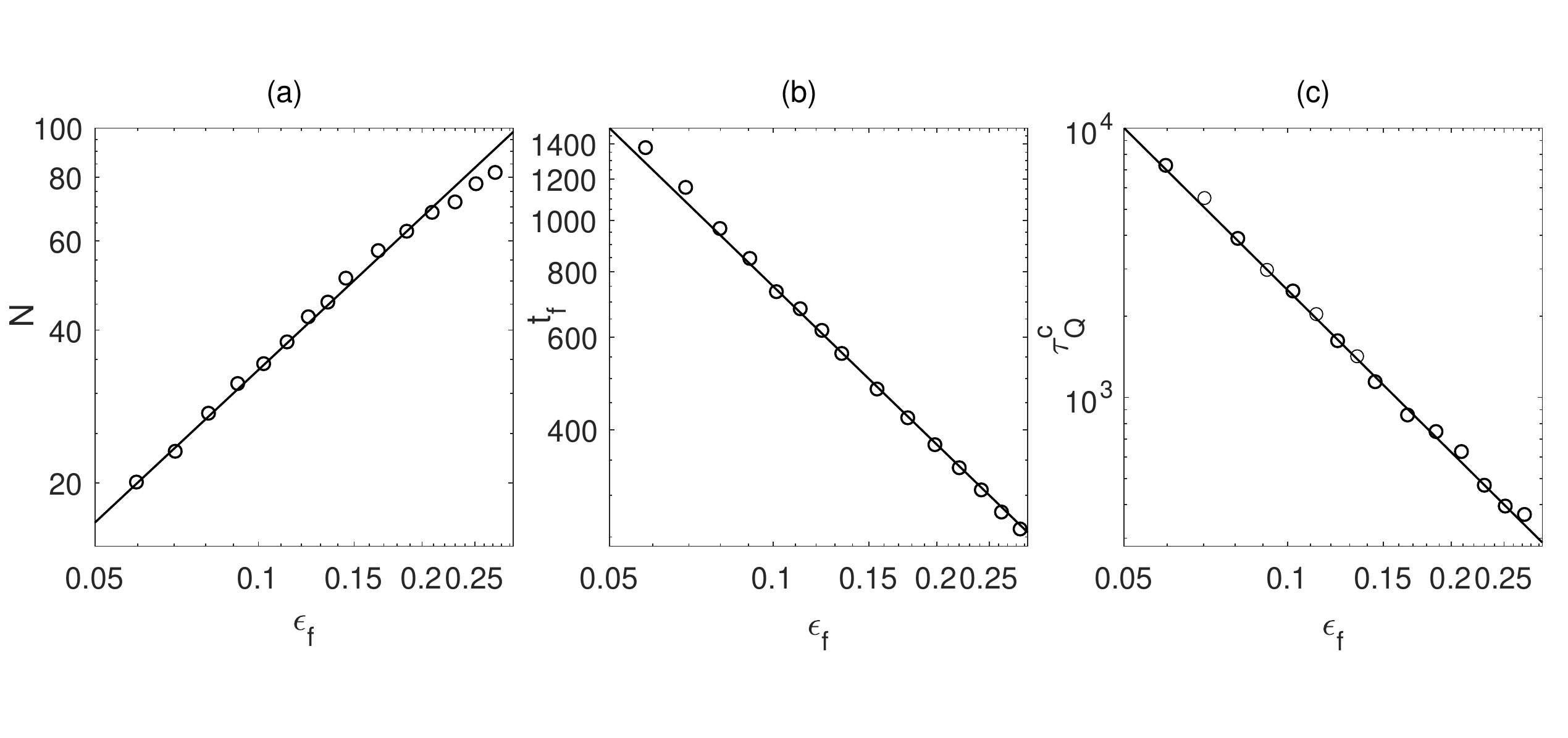}
    \caption{Universal dynamics beyond KZM after a fast quench of the electric field with an oscillating dynamic phase transition.  The panel (a) shows the linear scaling of the vortex number as a function of the final field intensity, i.e.,  $N\sim \epsilon_f^{1.003\pm0.0351}$.  The  panel (b) shows the dependence of the freeze-out time on $\epsilon_f$, with data fitted to $\hat{t}\sim \epsilon_f^{-1.042\pm 0.014}$. The  panel (c) shows the dependence of the critical quench time on $\epsilon_f$, with data fitted to $\tau_Q^c\sim \epsilon_f^{-1.985\pm 0.031}$.}
    \label{figure4}
\end{figure*}

\subsection{Kibble-Zurek Mechanism in slow and fast region}

According to our numerical results, in the slow quench region $\tau_Q > \tau_Q^c$, the number of vortices and quench rate satisfy the scaling $N \sim \tau_Q^{-1/2}$; see Fig. \ref{figure3}.
 The average critical exponents characterizing the order parameters in the oscillating field can be obtained by analyzing the quasinormal modes (QNMs) \cite{zeng2018universal}. The specific method is to use the perturbation equation and correlation function of the system to get the relationship between the correlation length and relaxation time and the external field, and average the results under the steady state condition, so as to get the average critical exponents. 
According to the calculation,
 the average correlation-length critical exponent is $\nu=1/2$ and the average dynamic critical exponent $z=2$.  Therefore, according to KZM, the topological defect density should follow a power law with exponent $-\frac{d\nu}{1+z\nu} = -1/2$. The validity of this prediction is confirmed by our results in the oscillating non-equilibrium phase transition.
The non-equilibrium KZM result we obtain does not depend on the time average, which means that the mean critical exponent can indeed describe the non-equilibrium phase transition driven by the periodic field.

However, in the fast quench region where $\tau_Q<\tau_Q^c$, the KZM power-law scaling is no longer satisfied, and the number of vortices remains constant, saturating at the value of the instantaneous quench scenario. 
Recently, it has been observed that in specific systems, even in this fast quench region where the KZM breaks down, a universal scaling behavior still holds \cite{chesler2015defect,xia2021kibble,zeng2023universal}. 
This theory relies on the fact that any realistic quench terminates at a finite value of $E_f$, which determines the lower limit of the relaxation time in the ordered phase.  Accordingly, the freeze-out time should be set by $\hat{t} \sim {\rm max} \{\tau[E(\hat{t}), \tau(E_f)]\}$, see Fig. \ref{figure0}. 
Then, for fast quenches with  $\tau(E_f) > \tau[E(\hat{t})]$, one finds a constant value of the freeze-out time, independent of $\tau_Q$, 
\begin{equation}
    \hat{t} \sim \tau(E_f) \sim \epsilon_f^{-z\nu} = \epsilon_f^{-1}.
\end{equation}
This explains the KZM breakdown and that the topological defect density saturates at a constant value which exhibits a distinct fast-quench universality.  As shown in the right panel of Fig. \ref{figure4}(a), the freeze-out time in the oscillating system agrees with the prediction $t_f \sim \epsilon_f^{-1}$. 
The relationship between the vortex density and the final field value can be derived from this scaling.  The average correlation length becomes $\hat{\xi}=\xi(E_f)$  and sets the average lengths scale of the domain.  The fast-quench vortex density  scales universally as
\begin{equation}
    n \sim \frac{1}{\xi(E_f)^d} \sim \epsilon_f^{d\nu}=\epsilon_f^1.
\end{equation}
%with $d\nu=1$.
 Figure \ref{figure4}(b) validates this result, showing that the universal scaling beyond the KZM  holds for the non-equilibrium phase transitions.  The non-equilibrium relaxation time does not affect the distribution of freeze-out time. 
 The critical quench rate can be defined by equating the time at which the quench ends at $E_f$, 
$t_f=\tau_Q^c(E_c-E_f)/E_c$. Then, $\tau_Q^c \sim \epsilon^{-(z\nu+1)}=\epsilon^{-2}$, as verified in Fig. \ref{figure4}(c).

\section{Summary and Discussion}  
Our study contributes to understanding non-equilibrium phase transitions driven by a periodic oscillatory field.
 We shed light on the rich dynamics inherent in these transitions by validating the Kibble-Zurek mechanism for slow quenches and uncovering the universality of rapid quenches in the oscillating dynamic phase transition. Moreover, we demonstrate the validity of the time-average critical exponent in periodic dynamic phase transitions by establishing non-equilibrium KZM scaling laws independent of the time average.
 The exploration of these phenomena not only deepens our understanding of fundamental nonequilibrium physics but also paves the way for potential applications in areas such as materials science, quantum computing, and energy technologies.  Our findings have significant implications for the applications of oscillating field-induced phase transitions in domains such as superconductivity, magnetism, and electric transport.  The validation of the Kibble-Zurek mechanism and the discovery of rapid-quench universality provide insights into the dynamics of these transitions, which can be leveraged to design and optimize materials with desired properties.  By tailoring the oscillatory field parameters and understanding the mechanisms driving the unconventional dynamics, we may achieve enhanced control over phase transitions, enabling advancements in superconducting technologies, magnetic devices, and systems for efficient electric and energy transport.
In addition, by varying the parameters of the oscillatory fields, such as frequency, amplitude, or waveform, it may be possible to manipulate the formation and evolution of topological defects, leading to intriguing possibilities for controlling and engineering novel materials and functionalities.

%\section*{Acknowledge}
{\it Acknowledgements.}
H.B. Z. acknowledges the support by the National Natural Science Foundation of China (under Grants No. 12275233), M. T. acknowledges the support by the
JSPS KAKENHI (under Grant No. JP20H01855) and W.C Y. acknowledges the support by JST, the establishment of university  fellowships towards the creation of science technology innovation (Under Grant No. JPMJFS2138). This research was funded in part by the Luxembourg National Research Fund (FNR), grant reference 17132060. For the purpose of open access, the authors have applied a Creative Commons Attribution 4.0 International (CC BY 4.0) license to any Author Accepted Manuscript version arising from this submission.

%\bibliography{reference}

\begin{thebibliography}{60}%
	\makeatletter
	\providecommand \@ifxundefined [1]{%
		\@ifx{#1\undefined}
	}%
	\providecommand \@ifnum [1]{%
		\ifnum #1\expandafter \@firstoftwo
		\else \expandafter \@secondoftwo
		\fi
	}%
	\providecommand \@ifx [1]{%
		\ifx #1\expandafter \@firstoftwo
		\else \expandafter \@secondoftwo
		\fi
	}%
	\providecommand \natexlab [1]{#1}%
	\providecommand \enquote  [1]{``#1''}%
	\providecommand \bibnamefont  [1]{#1}%
	\providecommand \bibfnamefont [1]{#1}%
	\providecommand \citenamefont [1]{#1}%
	\providecommand \href@noop [0]{\@secondoftwo}%
	\providecommand \href [0]{\begingroup \@sanitize@url \@href}%
	\providecommand \@href[1]{\@@startlink{#1}\@@href}%
	\providecommand \@@href[1]{\endgroup#1\@@endlink}%
	\providecommand \@sanitize@url [0]{\catcode `\\12\catcode `\$12\catcode
		`\&12\catcode `\#12\catcode `\^12\catcode `\_12\catcode `\%12\relax}%
	\providecommand \@@startlink[1]{}%
	\providecommand \@@endlink[0]{}%
	\providecommand \url  [0]{\begingroup\@sanitize@url \@url }%
	\providecommand \@url [1]{\endgroup\@href {#1}{\urlprefix }}%
	\providecommand \urlprefix  [0]{URL }%
	\providecommand \Eprint [0]{\href }%
	\providecommand \doibase [0]{https://doi.org/}%
	\providecommand \selectlanguage [0]{\@gobble}%
	\providecommand \bibinfo  [0]{\@secondoftwo}%
	\providecommand \bibfield  [0]{\@secondoftwo}%
	\providecommand \translation [1]{[#1]}%
	\providecommand \BibitemOpen [0]{}%
	\providecommand \bibitemStop [0]{}%
	\providecommand \bibitemNoStop [0]{.\EOS\space}%
	\providecommand \EOS [0]{\spacefactor3000\relax}%
	\providecommand \BibitemShut  [1]{\csname bibitem#1\endcsname}%
	\let\auto@bib@innerbib\@empty
	%</preamble>
	\bibitem [{\citenamefont {Kibble}(1976)}]{kibble1976topology}%
	\BibitemOpen
	\bibfield  {author} {\bibinfo {author} {\bibfnamefont {T.~W.~B.}\
			\bibnamefont {Kibble}},\ }\bibfield  {title} {\bibinfo {title} {Topology of
			cosmic domains and strings},\ }\href
	{https://doi.org/10.1088/0305-4470/9/8/029} {\bibfield  {journal} {\bibinfo
			{journal} {Journal of Physics A: Mathematical and General}\ }\textbf
		{\bibinfo {volume} {9}},\ \bibinfo {pages} {1387} (\bibinfo {year}
		{1976})}\BibitemShut {NoStop}%
	\bibitem [{\citenamefont {Zurek}(1985)}]{zurek1985cosmological}%
	\BibitemOpen
	\bibfield  {author} {\bibinfo {author} {\bibfnamefont {W.~H.}\ \bibnamefont
			{Zurek}},\ }\bibfield  {title} {\bibinfo {title} {Cosmological experiments in
			superfluid helium?},\ }\href {https://doi.org/10.1038/317505a0} {\bibfield
		{journal} {\bibinfo  {journal} {Nature}\ }\textbf {\bibinfo {volume} {317}},\
		\bibinfo {pages} {505–508} (\bibinfo {year} {1985})}\BibitemShut {NoStop}%
	\bibitem [{\citenamefont {del Campo}\ and\ \citenamefont
		{Zurek}(2014)}]{delcampo14}%
	\BibitemOpen
	\bibfield  {author} {\bibinfo {author} {\bibfnamefont {A.}~\bibnamefont {del
				Campo}}\ and\ \bibinfo {author} {\bibfnamefont {W.~H.}\ \bibnamefont
			{Zurek}},\ }\bibfield  {title} {\bibinfo {title} {Universality of phase
			transition dynamics: Topological defects from symmetry breaking},\ }\href
	{https://doi.org/10.1142/S0217751X1430018X} {\bibfield  {journal} {\bibinfo
			{journal} {International Journal of Modern Physics A}\ }\textbf {\bibinfo
			{volume} {29}},\ \bibinfo {pages} {1430018} (\bibinfo {year}
		{2014})}\BibitemShut {NoStop}%
	\bibitem [{\citenamefont {Chuang}\ \emph {et~al.}(1991)\citenamefont {Chuang},
		\citenamefont {Durrer}, \citenamefont {Turok},\ and\ \citenamefont
		{Yurke}}]{chuang1991cosmology}%
	\BibitemOpen
	\bibfield  {author} {\bibinfo {author} {\bibfnamefont {I.}~\bibnamefont
			{Chuang}}, \bibinfo {author} {\bibfnamefont {R.}~\bibnamefont {Durrer}},
		\bibinfo {author} {\bibfnamefont {N.}~\bibnamefont {Turok}},\ and\ \bibinfo
		{author} {\bibfnamefont {B.}~\bibnamefont {Yurke}},\ }\bibfield  {title}
	{\bibinfo {title} {Cosmology in the laboratory: Defect dynamics in liquid
			crystals},\ }\href {https://doi.org/10.1126/science.251.4999.1336} {\bibfield
		{journal} {\bibinfo  {journal} {Science}\ }\textbf {\bibinfo {volume}
			{251}},\ \bibinfo {pages} {1336} (\bibinfo {year} {1991})}\BibitemShut
	{NoStop}%
	\bibitem [{\citenamefont {Mielenz}\ \emph {et~al.}(2013)\citenamefont
		{Mielenz}, \citenamefont {Brox}, \citenamefont {Kahra}, \citenamefont
		{Leschhorn}, \citenamefont {Albert}, \citenamefont {Schaetz}, \citenamefont
		{Landa},\ and\ \citenamefont {Reznik}}]{mielenz2013trapping}%
	\BibitemOpen
	\bibfield  {author} {\bibinfo {author} {\bibfnamefont {M.}~\bibnamefont
			{Mielenz}}, \bibinfo {author} {\bibfnamefont {J.}~\bibnamefont {Brox}},
		\bibinfo {author} {\bibfnamefont {S.}~\bibnamefont {Kahra}}, \bibinfo
		{author} {\bibfnamefont {G.}~\bibnamefont {Leschhorn}}, \bibinfo {author}
		{\bibfnamefont {M.}~\bibnamefont {Albert}}, \bibinfo {author} {\bibfnamefont
			{T.}~\bibnamefont {Schaetz}}, \bibinfo {author} {\bibfnamefont
			{H.}~\bibnamefont {Landa}},\ and\ \bibinfo {author} {\bibfnamefont
			{B.}~\bibnamefont {Reznik}},\ }\bibfield  {title} {\bibinfo {title} {Trapping
			of topological-structural defects in coulomb crystals},\ }\href
	{https://doi.org/10.1103/PhysRevLett.110.133004} {\bibfield  {journal}
		{\bibinfo  {journal} {Phys. Rev. Lett.}\ }\textbf {\bibinfo {volume} {110}},\
		\bibinfo {pages} {133004} (\bibinfo {year} {2013})}\BibitemShut {NoStop}%
	\bibitem [{\citenamefont {Lamporesi}\ \emph {et~al.}(2013)\citenamefont
		{Lamporesi}, \citenamefont {Donadello}, \citenamefont {Serafini},
		\citenamefont {Dalfovo},\ and\ \citenamefont
		{Ferrari}}]{lamporesi2013spontaneous}%
	\BibitemOpen
	\bibfield  {author} {\bibinfo {author} {\bibfnamefont {G.}~\bibnamefont
			{Lamporesi}}, \bibinfo {author} {\bibfnamefont {S.}~\bibnamefont
			{Donadello}}, \bibinfo {author} {\bibfnamefont {S.}~\bibnamefont {Serafini}},
		\bibinfo {author} {\bibfnamefont {F.}~\bibnamefont {Dalfovo}},\ and\ \bibinfo
		{author} {\bibfnamefont {G.}~\bibnamefont {Ferrari}},\ }\bibfield  {title}
	{\bibinfo {title} {Spontaneous creation of kibble{\textendash}zurek solitons
			in a bose{\textendash}einstein condensate},\ }\href
	{https://doi.org/10.1038/nphys2734} {\bibfield  {journal} {\bibinfo
			{journal} {Nature Physics}\ }\textbf {\bibinfo {volume} {9}},\ \bibinfo
		{pages} {656} (\bibinfo {year} {2013})}\BibitemShut {NoStop}%
	\bibitem [{\citenamefont {Polkovnikov}\ \emph {et~al.}(2011)\citenamefont
		{Polkovnikov}, \citenamefont {Sengupta}, \citenamefont {Silva},\ and\
		\citenamefont {Vengalattore}}]{polkovnikov2011colloquium}%
	\BibitemOpen
	\bibfield  {author} {\bibinfo {author} {\bibfnamefont {A.}~\bibnamefont
			{Polkovnikov}}, \bibinfo {author} {\bibfnamefont {K.}~\bibnamefont
			{Sengupta}}, \bibinfo {author} {\bibfnamefont {A.}~\bibnamefont {Silva}},\
		and\ \bibinfo {author} {\bibfnamefont {M.}~\bibnamefont {Vengalattore}},\
	}\bibfield  {title} {\bibinfo {title} {Colloquium: Nonequilibrium dynamics of
			closed interacting quantum systems},\ }\href
	{https://doi.org/10.1103/RevModPhys.83.863} {\bibfield  {journal} {\bibinfo
			{journal} {Rev. Mod. Phys.}\ }\textbf {\bibinfo {volume} {83}},\ \bibinfo
		{pages} {863} (\bibinfo {year} {2011})}\BibitemShut {NoStop}%
	\bibitem [{\citenamefont {Weiler}\ \emph {et~al.}(2008)\citenamefont {Weiler},
		\citenamefont {Neely}, \citenamefont {Scherer}, \citenamefont {Bradley},
		\citenamefont {Davis},\ and\ \citenamefont
		{Anderson}}]{weiler2008spontaneous}%
	\BibitemOpen
	\bibfield  {author} {\bibinfo {author} {\bibfnamefont {C.~N.}\ \bibnamefont
			{Weiler}}, \bibinfo {author} {\bibfnamefont {T.~W.}\ \bibnamefont {Neely}},
		\bibinfo {author} {\bibfnamefont {D.~R.}\ \bibnamefont {Scherer}}, \bibinfo
		{author} {\bibfnamefont {A.~S.}\ \bibnamefont {Bradley}}, \bibinfo {author}
		{\bibfnamefont {M.~J.}\ \bibnamefont {Davis}},\ and\ \bibinfo {author}
		{\bibfnamefont {B.~P.}\ \bibnamefont {Anderson}},\ }\bibfield  {title}
	{\bibinfo {title} {Spontaneous vortices in the formation of
			bose{\textendash}einstein condensates},\ }\href
	{https://doi.org/10.1038/nature07334} {\bibfield  {journal} {\bibinfo
			{journal} {Nature}\ }\textbf {\bibinfo {volume} {455}},\ \bibinfo {pages}
		{948} (\bibinfo {year} {2008})}\BibitemShut {NoStop}%
	\bibitem [{\citenamefont {Dziarmaga}\ \emph {et~al.}(2008)\citenamefont
		{Dziarmaga}, \citenamefont {Meisner},\ and\ \citenamefont
		{Zurek}}]{dziarmaga2008winding}%
	\BibitemOpen
	\bibfield  {author} {\bibinfo {author} {\bibfnamefont {J.}~\bibnamefont
			{Dziarmaga}}, \bibinfo {author} {\bibfnamefont {J.}~\bibnamefont {Meisner}},\
		and\ \bibinfo {author} {\bibfnamefont {W.~H.}\ \bibnamefont {Zurek}},\
	}\bibfield  {title} {\bibinfo {title} {Winding up of the wave-function phase
			by an insulator-to-superfluid transition in a ring of coupled bose-einstein
			condensates},\ }\href {https://doi.org/10.1103/PhysRevLett.101.115701}
	{\bibfield  {journal} {\bibinfo  {journal} {Phys. Rev. Lett.}\ }\textbf
		{\bibinfo {volume} {101}},\ \bibinfo {pages} {115701} (\bibinfo {year}
		{2008})}\BibitemShut {NoStop}%
	\bibitem [{\citenamefont {Kolodrubetz}\ \emph {et~al.}(2012)\citenamefont
		{Kolodrubetz}, \citenamefont {Clark},\ and\ \citenamefont
		{Huse}}]{kolodrubetz2012nonequilibrium}%
	\BibitemOpen
	\bibfield  {author} {\bibinfo {author} {\bibfnamefont {M.}~\bibnamefont
			{Kolodrubetz}}, \bibinfo {author} {\bibfnamefont {B.~K.}\ \bibnamefont
			{Clark}},\ and\ \bibinfo {author} {\bibfnamefont {D.~A.}\ \bibnamefont
			{Huse}},\ }\bibfield  {title} {\bibinfo {title} {Nonequilibrium dynamic
			critical scaling of the quantum ising chain},\ }\href
	{https://doi.org/10.1103/PhysRevLett.109.015701} {\bibfield  {journal}
		{\bibinfo  {journal} {Phys. Rev. Lett.}\ }\textbf {\bibinfo {volume} {109}},\
		\bibinfo {pages} {015701} (\bibinfo {year} {2012})}\BibitemShut {NoStop}%
	\bibitem [{\citenamefont {Hohenberg}\ and\ \citenamefont
		{Halperin}(1977)}]{HohenbergHalperin77}%
	\BibitemOpen
	\bibfield  {author} {\bibinfo {author} {\bibfnamefont {P.~C.}\ \bibnamefont
			{Hohenberg}}\ and\ \bibinfo {author} {\bibfnamefont {B.~I.}\ \bibnamefont
			{Halperin}},\ }\bibfield  {title} {\bibinfo {title} {Theory of dynamic
			critical phenomena},\ }\href {https://doi.org/10.1103/RevModPhys.49.435}
	{\bibfield  {journal} {\bibinfo  {journal} {Rev. Mod. Phys.}\ }\textbf
		{\bibinfo {volume} {49}},\ \bibinfo {pages} {435} (\bibinfo {year}
		{1977})}\BibitemShut {NoStop}%
	\bibitem [{\citenamefont {Bowick}\ \emph {et~al.}(1994)\citenamefont {Bowick},
		\citenamefont {Chandar}, \citenamefont {Schiff},\ and\ \citenamefont
		{Srivastava}}]{bowick1994cosmological}%
	\BibitemOpen
	\bibfield  {author} {\bibinfo {author} {\bibfnamefont {M.~J.}\ \bibnamefont
			{Bowick}}, \bibinfo {author} {\bibfnamefont {L.}~\bibnamefont {Chandar}},
		\bibinfo {author} {\bibfnamefont {E.~A.}\ \bibnamefont {Schiff}},\ and\
		\bibinfo {author} {\bibfnamefont {A.~M.}\ \bibnamefont {Srivastava}},\
	}\bibfield  {title} {\bibinfo {title} {The cosmological kibble mechanism in
			the laboratory: String formation in liquid crystals},\ }\href
	{https://doi.org/10.1126/science.263.5149.943} {\bibfield  {journal}
		{\bibinfo  {journal} {Science}\ }\textbf {\bibinfo {volume} {263}},\ \bibinfo
		{pages} {943} (\bibinfo {year} {1994})}\BibitemShut {NoStop}%
	\bibitem [{\citenamefont {Hendry}\ \emph {et~al.}(1994)\citenamefont {Hendry},
		\citenamefont {Lawson}, \citenamefont {Lee}, \citenamefont {McClintock},\
		and\ \citenamefont {Williams}}]{hendry1994cosmological}%
	\BibitemOpen
	\bibfield  {author} {\bibinfo {author} {\bibfnamefont {P.~C.}\ \bibnamefont
			{Hendry}}, \bibinfo {author} {\bibfnamefont {N.~S.}\ \bibnamefont {Lawson}},
		\bibinfo {author} {\bibfnamefont {R.~A.~M.}\ \bibnamefont {Lee}}, \bibinfo
		{author} {\bibfnamefont {P.~V.~E.}\ \bibnamefont {McClintock}},\ and\
		\bibinfo {author} {\bibfnamefont {C.~D.~H.}\ \bibnamefont {Williams}},\
	}\bibfield  {title} {\bibinfo {title} {Generation of defects in superfluid
			4he as an analogue of the formation of cosmic strings},\ }\href
	{https://doi.org/10.1038/368315a0} {\bibfield  {journal} {\bibinfo  {journal}
			{Nature}\ }\textbf {\bibinfo {volume} {368}},\ \bibinfo {pages} {315}
		(\bibinfo {year} {1994})}\BibitemShut {NoStop}%
	\bibitem [{\citenamefont {Ruutu}\ \emph {et~al.}(1996)\citenamefont {Ruutu},
		\citenamefont {Eltsov}, \citenamefont {Gill}, \citenamefont {Kibble},
		\citenamefont {Krusius}, \citenamefont {Makhlin}, \citenamefont
		{Pla{\c{c}}ais}, \citenamefont {Volovik},\ and\ \citenamefont
		{Xu}}]{ruutu1996vortex}%
	\BibitemOpen
	\bibfield  {author} {\bibinfo {author} {\bibfnamefont {V.~M.~H.}\
			\bibnamefont {Ruutu}}, \bibinfo {author} {\bibfnamefont {V.~B.}\ \bibnamefont
			{Eltsov}}, \bibinfo {author} {\bibfnamefont {A.~J.}\ \bibnamefont {Gill}},
		\bibinfo {author} {\bibfnamefont {T.~W.~B.}\ \bibnamefont {Kibble}}, \bibinfo
		{author} {\bibfnamefont {M.}~\bibnamefont {Krusius}}, \bibinfo {author}
		{\bibfnamefont {Y.~G.}\ \bibnamefont {Makhlin}}, \bibinfo {author}
		{\bibfnamefont {B.}~\bibnamefont {Pla{\c{c}}ais}}, \bibinfo {author}
		{\bibfnamefont {G.~E.}\ \bibnamefont {Volovik}},\ and\ \bibinfo {author}
		{\bibfnamefont {W.}~\bibnamefont {Xu}},\ }\bibfield  {title} {\bibinfo
		{title} {Vortex formation in neutron-irradiated superfluid 3he as an analogue
			of cosmological defect formation},\ }\href {https://doi.org/10.1038/382334a0}
	{\bibfield  {journal} {\bibinfo  {journal} {Nature}\ }\textbf {\bibinfo
			{volume} {382}},\ \bibinfo {pages} {334} (\bibinfo {year}
		{1996})}\BibitemShut {NoStop}%
	\bibitem [{\citenamefont {Dodd}\ \emph {et~al.}(1998)\citenamefont {Dodd},
		\citenamefont {Hendry}, \citenamefont {Lawson}, \citenamefont {McClintock},\
		and\ \citenamefont {Williams}}]{dodd1998nonappearance}%
	\BibitemOpen
	\bibfield  {author} {\bibinfo {author} {\bibfnamefont {M.~E.}\ \bibnamefont
			{Dodd}}, \bibinfo {author} {\bibfnamefont {P.~C.}\ \bibnamefont {Hendry}},
		\bibinfo {author} {\bibfnamefont {N.~S.}\ \bibnamefont {Lawson}}, \bibinfo
		{author} {\bibfnamefont {P.~V.~E.}\ \bibnamefont {McClintock}},\ and\
		\bibinfo {author} {\bibfnamefont {C.~D.~H.}\ \bibnamefont {Williams}},\
	}\bibfield  {title} {\bibinfo {title} {Nonappearance of vortices in fast
			mechanical expansions of liquid ${}^{4}\mathrm{He}$ through the lambda
			transition},\ }\href {https://doi.org/10.1103/PhysRevLett.81.3703} {\bibfield
		{journal} {\bibinfo  {journal} {Phys. Rev. Lett.}\ }\textbf {\bibinfo
			{volume} {81}},\ \bibinfo {pages} {3703} (\bibinfo {year}
		{1998})}\BibitemShut {NoStop}%
	\bibitem [{\citenamefont {Sadler}\ \emph {et~al.}(2006)\citenamefont {Sadler},
		\citenamefont {Higbie}, \citenamefont {Leslie}, \citenamefont
		{Vengalattore},\ and\ \citenamefont {Stamper-Kurn}}]{sadler2006spontaneous}%
	\BibitemOpen
	\bibfield  {author} {\bibinfo {author} {\bibfnamefont {L.~E.}\ \bibnamefont
			{Sadler}}, \bibinfo {author} {\bibfnamefont {J.~M.}\ \bibnamefont {Higbie}},
		\bibinfo {author} {\bibfnamefont {S.~R.}\ \bibnamefont {Leslie}}, \bibinfo
		{author} {\bibfnamefont {M.}~\bibnamefont {Vengalattore}},\ and\ \bibinfo
		{author} {\bibfnamefont {D.~M.}\ \bibnamefont {Stamper-Kurn}},\ }\bibfield
	{title} {\bibinfo {title} {Spontaneous symmetry breaking in a quenched
			ferromagnetic spinor bose{\textendash}einstein condensate},\ }\href
	{https://doi.org/10.1038/nature05094} {\bibfield  {journal} {\bibinfo
			{journal} {Nature}\ }\textbf {\bibinfo {volume} {443}},\ \bibinfo {pages}
		{312} (\bibinfo {year} {2006})}\BibitemShut {NoStop}%
	\bibitem [{\citenamefont {Navon}\ \emph {et~al.}(2015)\citenamefont {Navon},
		\citenamefont {Gaunt}, \citenamefont {Smith},\ and\ \citenamefont
		{Hadzibabic}}]{navon2015critical}%
	\BibitemOpen
	\bibfield  {author} {\bibinfo {author} {\bibfnamefont {N.}~\bibnamefont
			{Navon}}, \bibinfo {author} {\bibfnamefont {A.~L.}\ \bibnamefont {Gaunt}},
		\bibinfo {author} {\bibfnamefont {R.~P.}\ \bibnamefont {Smith}},\ and\
		\bibinfo {author} {\bibfnamefont {Z.}~\bibnamefont {Hadzibabic}},\ }\bibfield
	{title} {\bibinfo {title} {Critical dynamics of spontaneous symmetry
			breaking in a homogeneous bose gas},\ }\href
	{https://doi.org/10.1126/science.1258676} {\bibfield  {journal} {\bibinfo
			{journal} {Science}\ }\textbf {\bibinfo {volume} {347}},\ \bibinfo {pages}
		{167} (\bibinfo {year} {2015})}\BibitemShut {NoStop}%
	\bibitem [{\citenamefont {Cui}\ \emph {et~al.}(2016)\citenamefont {Cui},
		\citenamefont {Huang}, \citenamefont {Wang}, \citenamefont {Cao},
		\citenamefont {Wang}, \citenamefont {Lv}, \citenamefont {Luo}, \citenamefont
		{del Campo}, \citenamefont {Han}, \citenamefont {Li},\ and\ \citenamefont
		{Guo}}]{Cui16}%
	\BibitemOpen
	\bibfield  {author} {\bibinfo {author} {\bibfnamefont {J.-M.}\ \bibnamefont
			{Cui}}, \bibinfo {author} {\bibfnamefont {Y.-F.}\ \bibnamefont {Huang}},
		\bibinfo {author} {\bibfnamefont {Z.}~\bibnamefont {Wang}}, \bibinfo {author}
		{\bibfnamefont {D.-Y.}\ \bibnamefont {Cao}}, \bibinfo {author} {\bibfnamefont
			{J.}~\bibnamefont {Wang}}, \bibinfo {author} {\bibfnamefont {W.-M.}\
			\bibnamefont {Lv}}, \bibinfo {author} {\bibfnamefont {L.}~\bibnamefont
			{Luo}}, \bibinfo {author} {\bibfnamefont {A.}~\bibnamefont {del Campo}},
		\bibinfo {author} {\bibfnamefont {Y.-J.}\ \bibnamefont {Han}}, \bibinfo
		{author} {\bibfnamefont {C.-F.}\ \bibnamefont {Li}},\ and\ \bibinfo {author}
		{\bibfnamefont {G.-C.}\ \bibnamefont {Guo}},\ }\bibfield  {title} {\bibinfo
		{title} {Experimental trapped-ion quantum simulation of the kibble-zurek
			dynamics in momentum space},\ }\href {https://doi.org/10.1038/srep33381}
	{\bibfield  {journal} {\bibinfo  {journal} {Scientific Reports}\ }\textbf
		{\bibinfo {volume} {6}},\ \bibinfo {pages} {33381} (\bibinfo {year}
		{2016})}\BibitemShut {NoStop}%
	\bibitem [{\citenamefont {Keesling}\ \emph {et~al.}(2019)\citenamefont
		{Keesling}, \citenamefont {Omran}, \citenamefont {Levine}, \citenamefont
		{Bernien}, \citenamefont {Pichler}, \citenamefont {Choi}, \citenamefont
		{Samajdar}, \citenamefont {Schwartz}, \citenamefont {Silvi}, \citenamefont
		{Sachdev}, \citenamefont {Zoller}, \citenamefont {Endres}, \citenamefont
		{Greiner}, \citenamefont {Vuleti{\'{c}}},\ and\ \citenamefont
		{Lukin}}]{Keesling19}%
	\BibitemOpen
	\bibfield  {author} {\bibinfo {author} {\bibfnamefont {A.}~\bibnamefont
			{Keesling}}, \bibinfo {author} {\bibfnamefont {A.}~\bibnamefont {Omran}},
		\bibinfo {author} {\bibfnamefont {H.}~\bibnamefont {Levine}}, \bibinfo
		{author} {\bibfnamefont {H.}~\bibnamefont {Bernien}}, \bibinfo {author}
		{\bibfnamefont {H.}~\bibnamefont {Pichler}}, \bibinfo {author} {\bibfnamefont
			{S.}~\bibnamefont {Choi}}, \bibinfo {author} {\bibfnamefont {R.}~\bibnamefont
			{Samajdar}}, \bibinfo {author} {\bibfnamefont {S.}~\bibnamefont {Schwartz}},
		\bibinfo {author} {\bibfnamefont {P.}~\bibnamefont {Silvi}}, \bibinfo
		{author} {\bibfnamefont {S.}~\bibnamefont {Sachdev}}, \bibinfo {author}
		{\bibfnamefont {P.}~\bibnamefont {Zoller}}, \bibinfo {author} {\bibfnamefont
			{M.}~\bibnamefont {Endres}}, \bibinfo {author} {\bibfnamefont
			{M.}~\bibnamefont {Greiner}}, \bibinfo {author} {\bibfnamefont
			{V.}~\bibnamefont {Vuleti{\'{c}}}},\ and\ \bibinfo {author} {\bibfnamefont
			{M.~D.}\ \bibnamefont {Lukin}},\ }\bibfield  {title} {\bibinfo {title}
		{Quantum kibble--zurek mechanism and critical dynamics on a programmable
			rydberg simulator},\ }\href {https://doi.org/10.1038/s41586-019-1070-1}
	{\bibfield  {journal} {\bibinfo  {journal} {Nature}\ }\textbf {\bibinfo
			{volume} {568}},\ \bibinfo {pages} {207} (\bibinfo {year}
		{2019})}\BibitemShut {NoStop}%
	\bibitem [{\citenamefont {Cui}\ \emph {et~al.}(2020)\citenamefont {Cui},
		\citenamefont {G{\'o}mez-Ruiz}, \citenamefont {Huang}, \citenamefont {Li},
		\citenamefont {Guo},\ and\ \citenamefont {del Campo}}]{Cui20}%
	\BibitemOpen
	\bibfield  {author} {\bibinfo {author} {\bibfnamefont {J.-M.}\ \bibnamefont
			{Cui}}, \bibinfo {author} {\bibfnamefont {F.~J.}\ \bibnamefont
			{G{\'o}mez-Ruiz}}, \bibinfo {author} {\bibfnamefont {Y.-F.}\ \bibnamefont
			{Huang}}, \bibinfo {author} {\bibfnamefont {C.-F.}\ \bibnamefont {Li}},
		\bibinfo {author} {\bibfnamefont {G.-C.}\ \bibnamefont {Guo}},\ and\ \bibinfo
		{author} {\bibfnamefont {A.}~\bibnamefont {del Campo}},\ }\bibfield  {title}
	{\bibinfo {title} {Experimentally testing quantum critical dynamics beyond
			the kibble--zurek mechanism},\ }\href
	{https://doi.org/10.1038/s42005-020-0306-6} {\bibfield  {journal} {\bibinfo
			{journal} {Communications Physics}\ }\textbf {\bibinfo {volume} {3}},\
		\bibinfo {pages} {44} (\bibinfo {year} {2020})}\BibitemShut {NoStop}%
	\bibitem [{\citenamefont {Bando}\ \emph {et~al.}(2020)\citenamefont {Bando},
		\citenamefont {Susa}, \citenamefont {Oshiyama}, \citenamefont {Shibata},
		\citenamefont {Ohzeki}, \citenamefont {G\'omez-Ruiz}, \citenamefont {Lidar},
		\citenamefont {Suzuki}, \citenamefont {del Campo},\ and\ \citenamefont
		{Nishimori}}]{Bando20}%
	\BibitemOpen
	\bibfield  {author} {\bibinfo {author} {\bibfnamefont {Y.}~\bibnamefont
			{Bando}}, \bibinfo {author} {\bibfnamefont {Y.}~\bibnamefont {Susa}},
		\bibinfo {author} {\bibfnamefont {H.}~\bibnamefont {Oshiyama}}, \bibinfo
		{author} {\bibfnamefont {N.}~\bibnamefont {Shibata}}, \bibinfo {author}
		{\bibfnamefont {M.}~\bibnamefont {Ohzeki}}, \bibinfo {author} {\bibfnamefont
			{F.~J.}\ \bibnamefont {G\'omez-Ruiz}}, \bibinfo {author} {\bibfnamefont
			{D.~A.}\ \bibnamefont {Lidar}}, \bibinfo {author} {\bibfnamefont
			{S.}~\bibnamefont {Suzuki}}, \bibinfo {author} {\bibfnamefont
			{A.}~\bibnamefont {del Campo}},\ and\ \bibinfo {author} {\bibfnamefont
			{H.}~\bibnamefont {Nishimori}},\ }\bibfield  {title} {\bibinfo {title}
		{Probing the universality of topological defect formation in a quantum
			annealer: Kibble-zurek mechanism and beyond},\ }\href
	{https://doi.org/10.1103/PhysRevResearch.2.033369} {\bibfield  {journal}
		{\bibinfo  {journal} {Phys. Rev. Research}\ }\textbf {\bibinfo {volume}
			{2}},\ \bibinfo {pages} {033369} (\bibinfo {year} {2020})}\BibitemShut
	{NoStop}%
	\bibitem [{\citenamefont {King}\ \emph {et~al.}(2022)\citenamefont {King},
		\citenamefont {Suzuki}, \citenamefont {Raymond}, \citenamefont {Zucca},
		\citenamefont {Lanting}, \citenamefont {Altomare}, \citenamefont {Berkley},
		\citenamefont {Ejtemaee}, \citenamefont {Hoskinson}, \citenamefont {Huang},
		\citenamefont {Ladizinsky}, \citenamefont {MacDonald}, \citenamefont
		{Marsden}, \citenamefont {Oh}, \citenamefont {Poulin-Lamarre}, \citenamefont
		{Reis}, \citenamefont {Rich}, \citenamefont {Sato}, \citenamefont
		{Whittaker}, \citenamefont {Yao}, \citenamefont {Harris}, \citenamefont
		{Lidar}, \citenamefont {Nishimori},\ and\ \citenamefont {Amin}}]{King22}%
	\BibitemOpen
	\bibfield  {author} {\bibinfo {author} {\bibfnamefont {A.~D.}\ \bibnamefont
			{King}}, \bibinfo {author} {\bibfnamefont {S.}~\bibnamefont {Suzuki}},
		\bibinfo {author} {\bibfnamefont {J.}~\bibnamefont {Raymond}}, \bibinfo
		{author} {\bibfnamefont {A.}~\bibnamefont {Zucca}}, \bibinfo {author}
		{\bibfnamefont {T.}~\bibnamefont {Lanting}}, \bibinfo {author} {\bibfnamefont
			{F.}~\bibnamefont {Altomare}}, \bibinfo {author} {\bibfnamefont {A.~J.}\
			\bibnamefont {Berkley}}, \bibinfo {author} {\bibfnamefont {S.}~\bibnamefont
			{Ejtemaee}}, \bibinfo {author} {\bibfnamefont {E.}~\bibnamefont {Hoskinson}},
		\bibinfo {author} {\bibfnamefont {S.}~\bibnamefont {Huang}}, \bibinfo
		{author} {\bibfnamefont {E.}~\bibnamefont {Ladizinsky}}, \bibinfo {author}
		{\bibfnamefont {A.~J.~R.}\ \bibnamefont {MacDonald}}, \bibinfo {author}
		{\bibfnamefont {G.}~\bibnamefont {Marsden}}, \bibinfo {author} {\bibfnamefont
			{T.}~\bibnamefont {Oh}}, \bibinfo {author} {\bibfnamefont {G.}~\bibnamefont
			{Poulin-Lamarre}}, \bibinfo {author} {\bibfnamefont {M.}~\bibnamefont
			{Reis}}, \bibinfo {author} {\bibfnamefont {C.}~\bibnamefont {Rich}}, \bibinfo
		{author} {\bibfnamefont {Y.}~\bibnamefont {Sato}}, \bibinfo {author}
		{\bibfnamefont {J.~D.}\ \bibnamefont {Whittaker}}, \bibinfo {author}
		{\bibfnamefont {J.}~\bibnamefont {Yao}}, \bibinfo {author} {\bibfnamefont
			{R.}~\bibnamefont {Harris}}, \bibinfo {author} {\bibfnamefont {D.~A.}\
			\bibnamefont {Lidar}}, \bibinfo {author} {\bibfnamefont {H.}~\bibnamefont
			{Nishimori}},\ and\ \bibinfo {author} {\bibfnamefont {M.~H.}\ \bibnamefont
			{Amin}},\ }\bibfield  {title} {\bibinfo {title} {Coherent quantum annealing
			in a programmable 2,000{\thinspace}qubit ising chain},\ }\href
	{https://doi.org/10.1038/s41567-022-01741-6} {\bibfield  {journal} {\bibinfo
			{journal} {Nature Physics}\ }\textbf {\bibinfo {volume} {18}},\ \bibinfo
		{pages} {1324} (\bibinfo {year} {2022})}\BibitemShut {NoStop}%
	\bibitem [{\citenamefont {Du}\ \emph {et~al.}(2023)\citenamefont {Du},
		\citenamefont {Fang}, \citenamefont {Won}, \citenamefont {De}, \citenamefont
		{Huang}, \citenamefont {Xu}, \citenamefont {You}, \citenamefont
		{G{\'o}mez-Ruiz}, \citenamefont {del Campo},\ and\ \citenamefont
		{Cheong}}]{Du2023}%
	\BibitemOpen
	\bibfield  {author} {\bibinfo {author} {\bibfnamefont {K.}~\bibnamefont
			{Du}}, \bibinfo {author} {\bibfnamefont {X.}~\bibnamefont {Fang}}, \bibinfo
		{author} {\bibfnamefont {C.}~\bibnamefont {Won}}, \bibinfo {author}
		{\bibfnamefont {C.}~\bibnamefont {De}}, \bibinfo {author} {\bibfnamefont
			{F.-T.}\ \bibnamefont {Huang}}, \bibinfo {author} {\bibfnamefont
			{W.}~\bibnamefont {Xu}}, \bibinfo {author} {\bibfnamefont {H.}~\bibnamefont
			{You}}, \bibinfo {author} {\bibfnamefont {F.~J.}\ \bibnamefont
			{G{\'o}mez-Ruiz}}, \bibinfo {author} {\bibfnamefont {A.}~\bibnamefont {del
				Campo}},\ and\ \bibinfo {author} {\bibfnamefont {S.-W.}\ \bibnamefont
			{Cheong}},\ }\bibfield  {title} {\bibinfo {title} {Kibble--zurek mechanism of
			ising domains},\ }\href {https://doi.org/10.1038/s41567-023-02112-5}
	{\bibfield  {journal} {\bibinfo  {journal} {Nature Physics}\ }\textbf
		{\bibinfo {volume} {19}},\ \bibinfo {pages} {1495} (\bibinfo {year}
		{2023})}\BibitemShut {NoStop}%
	\bibitem [{\citenamefont {Ducci}\ \emph {et~al.}(1999)\citenamefont {Ducci},
		\citenamefont {Ramazza}, \citenamefont {Gonz\'alez-Vi\~nas},\ and\
		\citenamefont {Arecchi}}]{Ducci99}%
	\BibitemOpen
	\bibfield  {author} {\bibinfo {author} {\bibfnamefont {S.}~\bibnamefont
			{Ducci}}, \bibinfo {author} {\bibfnamefont {P.~L.}\ \bibnamefont {Ramazza}},
		\bibinfo {author} {\bibfnamefont {W.}~\bibnamefont {Gonz\'alez-Vi\~nas}},\
		and\ \bibinfo {author} {\bibfnamefont {F.~T.}\ \bibnamefont {Arecchi}},\
	}\bibfield  {title} {\bibinfo {title} {Order parameter fragmentation after a
			symmetry-breaking transition},\ }\href
	{https://doi.org/10.1103/PhysRevLett.83.5210} {\bibfield  {journal} {\bibinfo
			{journal} {Phys. Rev. Lett.}\ }\textbf {\bibinfo {volume} {83}},\ \bibinfo
		{pages} {5210} (\bibinfo {year} {1999})}\BibitemShut {NoStop}%
	\bibitem [{\citenamefont {Casado}\ \emph {et~al.}(2001)\citenamefont {Casado},
		\citenamefont {Gonz\'alez-Vi\~nas}, \citenamefont {Mancini},\ and\
		\citenamefont {Boccaletti}}]{Casado01}%
	\BibitemOpen
	\bibfield  {author} {\bibinfo {author} {\bibfnamefont {S.}~\bibnamefont
			{Casado}}, \bibinfo {author} {\bibfnamefont {W.}~\bibnamefont
			{Gonz\'alez-Vi\~nas}}, \bibinfo {author} {\bibfnamefont {H.}~\bibnamefont
			{Mancini}},\ and\ \bibinfo {author} {\bibfnamefont {S.}~\bibnamefont
			{Boccaletti}},\ }\bibfield  {title} {\bibinfo {title} {Topological defects
			after a quench in a {B\'enard-Marangoni} convection system},\ }\href
	{https://doi.org/10.1103/PhysRevE.63.057301} {\bibfield  {journal} {\bibinfo
			{journal} {Phys. Rev. E}\ }\textbf {\bibinfo {volume} {63}},\ \bibinfo
		{pages} {057301} (\bibinfo {year} {2001})}\BibitemShut {NoStop}%
	\bibitem [{\citenamefont {Casado}\ \emph {et~al.}(2006)\citenamefont {Casado},
		\citenamefont {Gonz\'alez-Vi\~nas},\ and\ \citenamefont
		{Mancini}}]{Casado06}%
	\BibitemOpen
	\bibfield  {author} {\bibinfo {author} {\bibfnamefont {S.}~\bibnamefont
			{Casado}}, \bibinfo {author} {\bibfnamefont {W.}~\bibnamefont
			{Gonz\'alez-Vi\~nas}},\ and\ \bibinfo {author} {\bibfnamefont
			{H.}~\bibnamefont {Mancini}},\ }\bibfield  {title} {\bibinfo {title} {Testing
			the {Kibble-Zurek} mechanism in {Rayleigh-B\'enard} convection},\ }\href
	{https://doi.org/10.1103/PhysRevE.74.047101} {\bibfield  {journal} {\bibinfo
			{journal} {Phys. Rev. E}\ }\textbf {\bibinfo {volume} {74}},\ \bibinfo
		{pages} {047101} (\bibinfo {year} {2006})}\BibitemShut {NoStop}%
	\bibitem [{\citenamefont {Casado}\ \emph {et~al.}(2007)\citenamefont {Casado},
		\citenamefont {Gonz{\' a}lez-Vi{\~ n}as}, \citenamefont {Boccaletti},
		\citenamefont {Ramazza},\ and\ \citenamefont {Mancini}}]{Casado07}%
	\BibitemOpen
	\bibfield  {author} {\bibinfo {author} {\bibfnamefont {S.}~\bibnamefont
			{Casado}}, \bibinfo {author} {\bibfnamefont {W.}~\bibnamefont {Gonz{\'
					a}lez-Vi{\~ n}as}}, \bibinfo {author} {\bibfnamefont {S.}~\bibnamefont
			{Boccaletti}}, \bibinfo {author} {\bibfnamefont {P.~L.}\ \bibnamefont
			{Ramazza}},\ and\ \bibinfo {author} {\bibfnamefont {H.}~\bibnamefont
			{Mancini}},\ }\bibfield  {title} {\bibinfo {title} {The birth of defects in
			pattern formation: Testing of the {Kibble-Zurek} mechanism},\ }\href
	{https://doi.org/10.1140/epjst/e2007-00171-2} {\bibfield  {journal} {\bibinfo
			{journal} {Eur. Phys. J. Special Topics}\ }\textbf {\bibinfo {volume}
			{146}},\ \bibinfo {pages} {87} (\bibinfo {year} {2007})}\BibitemShut
	{NoStop}%
	\bibitem [{\citenamefont {Miranda}\ \emph {et~al.}(2012)\citenamefont
		{Miranda}, \citenamefont {Burguete}, \citenamefont {Gonz{\' a}lez-Vi{\~
				n}as},\ and\ \citenamefont {Mancini}}]{Miranda12}%
	\BibitemOpen
	\bibfield  {author} {\bibinfo {author} {\bibfnamefont {M.~A.}\ \bibnamefont
			{Miranda}}, \bibinfo {author} {\bibfnamefont {J.}~\bibnamefont {Burguete}},
		\bibinfo {author} {\bibfnamefont {W.}~\bibnamefont {Gonz{\' a}lez-Vi{\~
					n}as}},\ and\ \bibinfo {author} {\bibfnamefont {H.}~\bibnamefont {Mancini}},\
	}\bibfield  {title} {\bibinfo {title} {Exploring the {Kibble-Zurek} mechanism
			in a secondary bifurcation},\ }\href
	{https://doi.org/10.1142/S0218127412501659} {\bibfield  {journal} {\bibinfo
			{journal} {Int. J. Bifurcation Chaos}\ }\textbf {\bibinfo {volume} {22}},\
		\bibinfo {pages} {1250165} (\bibinfo {year} {2012})}\BibitemShut {NoStop}%
	\bibitem [{\citenamefont {Miranda}\ \emph {et~al.}(2013)\citenamefont
		{Miranda}, \citenamefont {Burguete}, \citenamefont {Mancini},\ and\
		\citenamefont {Gonz\'alez-Vi\~nas}}]{Miranda13}%
	\BibitemOpen
	\bibfield  {author} {\bibinfo {author} {\bibfnamefont {M.~A.}\ \bibnamefont
			{Miranda}}, \bibinfo {author} {\bibfnamefont {J.}~\bibnamefont {Burguete}},
		\bibinfo {author} {\bibfnamefont {H.}~\bibnamefont {Mancini}},\ and\ \bibinfo
		{author} {\bibfnamefont {W.}~\bibnamefont {Gonz\'alez-Vi\~nas}},\ }\bibfield
	{title} {\bibinfo {title} {Frozen dynamics and synchronization through a
			secondary symmetry-breaking bifurcation},\ }\href
	{https://doi.org/10.1103/PhysRevE.87.032902} {\bibfield  {journal} {\bibinfo
			{journal} {Phys. Rev. E}\ }\textbf {\bibinfo {volume} {87}},\ \bibinfo
		{pages} {032902} (\bibinfo {year} {2013})}\BibitemShut {NoStop}%
	\bibitem [{\citenamefont {Zamora}\ \emph {et~al.}(2020)\citenamefont {Zamora},
		\citenamefont {Dagvadorj}, \citenamefont {Comaron}, \citenamefont
		{Carusotto}, \citenamefont {Proukakis},\ and\ \citenamefont
		{Szyma\ifmmode~\acute{n}\else \'{n}\fi{}ska}}]{zamora2020kibble}%
	\BibitemOpen
	\bibfield  {author} {\bibinfo {author} {\bibfnamefont {A.}~\bibnamefont
			{Zamora}}, \bibinfo {author} {\bibfnamefont {G.}~\bibnamefont {Dagvadorj}},
		\bibinfo {author} {\bibfnamefont {P.}~\bibnamefont {Comaron}}, \bibinfo
		{author} {\bibfnamefont {I.}~\bibnamefont {Carusotto}}, \bibinfo {author}
		{\bibfnamefont {N.~P.}\ \bibnamefont {Proukakis}},\ and\ \bibinfo {author}
		{\bibfnamefont {M.~H.}\ \bibnamefont {Szyma\ifmmode~\acute{n}\else
				\'{n}\fi{}ska}},\ }\bibfield  {title} {\bibinfo {title} {Kibble-zurek
			mechanism in driven dissipative systems crossing a nonequilibrium phase
			transition},\ }\href {https://doi.org/10.1103/PhysRevLett.125.095301}
	{\bibfield  {journal} {\bibinfo  {journal} {Phys. Rev. Lett.}\ }\textbf
		{\bibinfo {volume} {125}},\ \bibinfo {pages} {095301} (\bibinfo {year}
		{2020})}\BibitemShut {NoStop}%
	\bibitem [{\citenamefont {Reichhardt}\ \emph {et~al.}(2022)\citenamefont
		{Reichhardt}, \citenamefont {del Campo},\ and\ \citenamefont
		{Reichhardt}}]{reichhardt2022kibble}%
	\BibitemOpen
	\bibfield  {author} {\bibinfo {author} {\bibfnamefont {C.~J.~O.}\
			\bibnamefont {Reichhardt}}, \bibinfo {author} {\bibfnamefont
			{A.}~\bibnamefont {del Campo}},\ and\ \bibinfo {author} {\bibfnamefont
			{C.}~\bibnamefont {Reichhardt}},\ }\bibfield  {title} {\bibinfo {title}
		{Kibble-zurek mechanism for nonequilibrium phase transitions in driven
			systems with quenched disorder},\ }\href
	{https://doi.org/10.1038/s42005-022-00952-w} {\bibfield  {journal} {\bibinfo
			{journal} {Communications Physics}\ }\textbf {\bibinfo {volume} {5}},\
		\bibinfo {pages} {173} (\bibinfo {year} {2022})}\BibitemShut {NoStop}%
	\bibitem [{\citenamefont {Maegochi}\ \emph {et~al.}(2022)\citenamefont
		{Maegochi}, \citenamefont {Ienaga},\ and\ \citenamefont
		{Okuma}}]{maegochi2022kibble}%
	\BibitemOpen
	\bibfield  {author} {\bibinfo {author} {\bibfnamefont {S.}~\bibnamefont
			{Maegochi}}, \bibinfo {author} {\bibfnamefont {K.}~\bibnamefont {Ienaga}},\
		and\ \bibinfo {author} {\bibfnamefont {S.}~\bibnamefont {Okuma}},\ }\bibfield
	{title} {\bibinfo {title} {Kibble-zurek mechanism for dynamical ordering in
			a driven vortex system},\ }\href
	{https://doi.org/10.1103/PhysRevLett.129.227001} {\bibfield  {journal}
		{\bibinfo  {journal} {Phys. Rev. Lett.}\ }\textbf {\bibinfo {volume} {129}},\
		\bibinfo {pages} {227001} (\bibinfo {year} {2022})}\BibitemShut {NoStop}%
	\bibitem [{\citenamefont {Sides}\ \emph {et~al.}(1998)\citenamefont {Sides},
		\citenamefont {Rikvold},\ and\ \citenamefont {Novotny}}]{sides1998kinetic}%
	\BibitemOpen
	\bibfield  {author} {\bibinfo {author} {\bibfnamefont {S.~W.}\ \bibnamefont
			{Sides}}, \bibinfo {author} {\bibfnamefont {P.~A.}\ \bibnamefont {Rikvold}},\
		and\ \bibinfo {author} {\bibfnamefont {M.~A.}\ \bibnamefont {Novotny}},\
	}\bibfield  {title} {\bibinfo {title} {Kinetic ising model in an oscillating
			field: Finite-size scaling at the dynamic phase transition},\ }\href
	{https://doi.org/10.1103/PhysRevLett.81.834} {\bibfield  {journal} {\bibinfo
			{journal} {Phys. Rev. Lett.}\ }\textbf {\bibinfo {volume} {81}},\ \bibinfo
		{pages} {834} (\bibinfo {year} {1998})}\BibitemShut {NoStop}%
	\bibitem [{\citenamefont {Tom\'e}\ and\ \citenamefont
		{de~Oliveira}(1990)}]{tome1990dynamic}%
	\BibitemOpen
	\bibfield  {author} {\bibinfo {author} {\bibfnamefont {T.}~\bibnamefont
			{Tom\'e}}\ and\ \bibinfo {author} {\bibfnamefont {M.~J.}\ \bibnamefont
			{de~Oliveira}},\ }\bibfield  {title} {\bibinfo {title} {Dynamic phase
			transition in the kinetic ising model under a time-dependent oscillating
			field},\ }\href {https://doi.org/10.1103/PhysRevA.41.4251} {\bibfield
		{journal} {\bibinfo  {journal} {Phys. Rev. A}\ }\textbf {\bibinfo {volume}
			{41}},\ \bibinfo {pages} {4251} (\bibinfo {year} {1990})}\BibitemShut
	{NoStop}%
	\bibitem [{\citenamefont {Y\"uksel}\ \emph {et~al.}(2012)\citenamefont
		{Y\"uksel}, \citenamefont {Vatansever}, \citenamefont {Ak\ifmmode \imath
			\else \i \fi{}nc\ifmmode \imath \else~\i \fi{}},\ and\ \citenamefont
		{Polat}}]{yuksel2012nonequilibrium}%
	\BibitemOpen
	\bibfield  {author} {\bibinfo {author} {\bibfnamefont {Y.}~\bibnamefont
			{Y\"uksel}}, \bibinfo {author} {\bibfnamefont {E.}~\bibnamefont
			{Vatansever}}, \bibinfo {author} {\bibfnamefont {U.}~\bibnamefont {Ak\ifmmode
				\imath \else \i \fi{}nc\ifmmode \imath \else~\i \fi{}}},\ and\ \bibinfo
		{author} {\bibfnamefont {H.}~\bibnamefont {Polat}},\ }\bibfield  {title}
	{\bibinfo {title} {Nonequilibrium phase transitions and stationary-state
			solutions of a three-dimensional random-field ising model under a
			time-dependent periodic external field},\ }\href
	{https://doi.org/10.1103/PhysRevE.85.051123} {\bibfield  {journal} {\bibinfo
			{journal} {Phys. Rev. E}\ }\textbf {\bibinfo {volume} {85}},\ \bibinfo
		{pages} {051123} (\bibinfo {year} {2012})}\BibitemShut {NoStop}%
	\bibitem [{\citenamefont {Collado}\ \emph {et~al.}(2023)\citenamefont
		{Collado}, \citenamefont {Usaj}, \citenamefont {Balseiro}, \citenamefont
		{Zanette},\ and\ \citenamefont {Lorenzana}}]{PhysRevResearch.5.023014}%
	\BibitemOpen
	\bibfield  {author} {\bibinfo {author} {\bibfnamefont {H.~P.~O.}\
			\bibnamefont {Collado}}, \bibinfo {author} {\bibfnamefont {G.}~\bibnamefont
			{Usaj}}, \bibinfo {author} {\bibfnamefont {C.~A.}\ \bibnamefont {Balseiro}},
		\bibinfo {author} {\bibfnamefont {D.~H.}\ \bibnamefont {Zanette}},\ and\
		\bibinfo {author} {\bibfnamefont {J.}~\bibnamefont {Lorenzana}},\ }\bibfield
	{title} {\bibinfo {title} {Dynamical phase transitions in periodically driven
			bardeen-cooper-schrieffer systems},\ }\href
	{https://doi.org/10.1103/PhysRevResearch.5.023014} {\bibfield  {journal}
		{\bibinfo  {journal} {Phys. Rev. Res.}\ }\textbf {\bibinfo {volume} {5}},\
		\bibinfo {pages} {023014} (\bibinfo {year} {2023})}\BibitemShut {NoStop}%
	\bibitem [{\citenamefont {Eckardt}\ \emph {et~al.}(2005)\citenamefont
		{Eckardt}, \citenamefont {Weiss},\ and\ \citenamefont
		{Holthaus}}]{PhysRevLett.95.260404}%
	\BibitemOpen
	\bibfield  {author} {\bibinfo {author} {\bibfnamefont {A.}~\bibnamefont
			{Eckardt}}, \bibinfo {author} {\bibfnamefont {C.}~\bibnamefont {Weiss}},\
		and\ \bibinfo {author} {\bibfnamefont {M.}~\bibnamefont {Holthaus}},\
	}\bibfield  {title} {\bibinfo {title} {Superfluid-insulator transition in a
			periodically driven optical lattice},\ }\href
	{https://doi.org/10.1103/PhysRevLett.95.260404} {\bibfield  {journal}
		{\bibinfo  {journal} {Phys. Rev. Lett.}\ }\textbf {\bibinfo {volume} {95}},\
		\bibinfo {pages} {260404} (\bibinfo {year} {2005})}\BibitemShut {NoStop}%
	\bibitem [{\citenamefont {Reichhardt}\ \emph {et~al.}(2023)\citenamefont
		{Reichhardt}, \citenamefont {Regev}, \citenamefont {Dahmen}, \citenamefont
		{Okuma},\ and\ \citenamefont {Reichhardt}}]{PhysRevResearch.5.021001}%
	\BibitemOpen
	\bibfield  {author} {\bibinfo {author} {\bibfnamefont {C.}~\bibnamefont
			{Reichhardt}}, \bibinfo {author} {\bibfnamefont {I.}~\bibnamefont {Regev}},
		\bibinfo {author} {\bibfnamefont {K.}~\bibnamefont {Dahmen}}, \bibinfo
		{author} {\bibfnamefont {S.}~\bibnamefont {Okuma}},\ and\ \bibinfo {author}
		{\bibfnamefont {C.~J.~O.}\ \bibnamefont {Reichhardt}},\ }\bibfield  {title}
	{\bibinfo {title} {Reversible to irreversible transitions in periodic driven
			many-body systems and future directions for classical and quantum systems},\
	}\href {https://doi.org/10.1103/PhysRevResearch.5.021001} {\bibfield
		{journal} {\bibinfo  {journal} {Phys. Rev. Res.}\ }\textbf {\bibinfo {volume}
			{5}},\ \bibinfo {pages} {021001} (\bibinfo {year} {2023})}\BibitemShut
	{NoStop}%
	\bibitem [{\citenamefont {Zeng}\ and\ \citenamefont
		{Zhang}(2018)}]{zeng2018universal}%
	\BibitemOpen
	\bibfield  {author} {\bibinfo {author} {\bibfnamefont {H.-B.}\ \bibnamefont
			{Zeng}}\ and\ \bibinfo {author} {\bibfnamefont {H.-Q.}\ \bibnamefont
			{Zhang}},\ }\bibfield  {title} {\bibinfo {title} {Universal critical
			exponents of nonequilibrium phase transitions from holography},\ }\href
	{https://doi.org/10.1103/PhysRevD.98.106024} {\bibfield  {journal} {\bibinfo
			{journal} {Phys. Rev. D}\ }\textbf {\bibinfo {volume} {98}},\ \bibinfo
		{pages} {106024} (\bibinfo {year} {2018})}\BibitemShut {NoStop}%
	\bibitem [{\citenamefont {Holczer}\ \emph {et~al.}(1991)\citenamefont
		{Holczer}, \citenamefont {Klein}, \citenamefont {Gr\"uner}, \citenamefont
		{Thompson}, \citenamefont {Diederich},\ and\ \citenamefont
		{Whetten}}]{holczer1991critical}%
	\BibitemOpen
	\bibfield  {author} {\bibinfo {author} {\bibfnamefont {K.}~\bibnamefont
			{Holczer}}, \bibinfo {author} {\bibfnamefont {O.}~\bibnamefont {Klein}},
		\bibinfo {author} {\bibfnamefont {G.}~\bibnamefont {Gr\"uner}}, \bibinfo
		{author} {\bibfnamefont {J.~D.}\ \bibnamefont {Thompson}}, \bibinfo {author}
		{\bibfnamefont {F.}~\bibnamefont {Diederich}},\ and\ \bibinfo {author}
		{\bibfnamefont {R.~L.}\ \bibnamefont {Whetten}},\ }\bibfield  {title}
	{\bibinfo {title} {Critical magnetic fields in the superconducting state of
			${\mathrm{k}}_{3}$${\mathrm{c}}_{60}$},\ }\href
	{https://doi.org/10.1103/PhysRevLett.67.271} {\bibfield  {journal} {\bibinfo
			{journal} {Phys. Rev. Lett.}\ }\textbf {\bibinfo {volume} {67}},\ \bibinfo
		{pages} {271} (\bibinfo {year} {1991})}\BibitemShut {NoStop}%
	\bibitem [{\citenamefont {Langer}\ and\ \citenamefont
		{Fisher}(1967)}]{langer1967intrinsic}%
	\BibitemOpen
	\bibfield  {author} {\bibinfo {author} {\bibfnamefont {J.~S.}\ \bibnamefont
			{Langer}}\ and\ \bibinfo {author} {\bibfnamefont {M.~E.}\ \bibnamefont
			{Fisher}},\ }\bibfield  {title} {\bibinfo {title} {Intrinsic critical
			velocity of a superfluid},\ }\href
	{https://doi.org/10.1103/PhysRevLett.19.560} {\bibfield  {journal} {\bibinfo
			{journal} {Phys. Rev. Lett.}\ }\textbf {\bibinfo {volume} {19}},\ \bibinfo
		{pages} {560} (\bibinfo {year} {1967})}\BibitemShut {NoStop}%
	\bibitem [{\citenamefont {Herzog}\ \emph {et~al.}(2009)\citenamefont {Herzog},
		\citenamefont {Kovtun},\ and\ \citenamefont {Son}}]{herzog2009holographic}%
	\BibitemOpen
	\bibfield  {author} {\bibinfo {author} {\bibfnamefont {C.~P.}\ \bibnamefont
			{Herzog}}, \bibinfo {author} {\bibfnamefont {P.~K.}\ \bibnamefont {Kovtun}},\
		and\ \bibinfo {author} {\bibfnamefont {D.~T.}\ \bibnamefont {Son}},\
	}\bibfield  {title} {\bibinfo {title} {Holographic model of superfluidity},\
	}\href {https://doi.org/10.1103/PhysRevD.79.066002} {\bibfield  {journal}
		{\bibinfo  {journal} {Phys. Rev. D}\ }\textbf {\bibinfo {volume} {79}},\
		\bibinfo {pages} {066002} (\bibinfo {year} {2009})}\BibitemShut {NoStop}%
	\bibitem [{\citenamefont {Nakano}\ and\ \citenamefont
		{Wen}(2008)}]{nakano2008critical}%
	\BibitemOpen
	\bibfield  {author} {\bibinfo {author} {\bibfnamefont {E.}~\bibnamefont
			{Nakano}}\ and\ \bibinfo {author} {\bibfnamefont {W.-Y.}\ \bibnamefont
			{Wen}},\ }\bibfield  {title} {\bibinfo {title} {Critical magnetic field in a
			holographic superconductor},\ }\href
	{https://doi.org/10.1103/PhysRevD.78.046004} {\bibfield  {journal} {\bibinfo
			{journal} {Phys. Rev. D}\ }\textbf {\bibinfo {volume} {78}},\ \bibinfo
		{pages} {046004} (\bibinfo {year} {2008})}\BibitemShut {NoStop}%
	\bibitem [{\citenamefont {Yang}\ \emph {et~al.}(2023)\citenamefont {Yang},
		\citenamefont {Xia}, \citenamefont {Zeng}, \citenamefont {Tsubota},\ and\
		\citenamefont {Zaanen}}]{yang2023motion}%
	\BibitemOpen
	\bibfield  {author} {\bibinfo {author} {\bibfnamefont {W.-C.}\ \bibnamefont
			{Yang}}, \bibinfo {author} {\bibfnamefont {C.-Y.}\ \bibnamefont {Xia}},
		\bibinfo {author} {\bibfnamefont {H.-B.}\ \bibnamefont {Zeng}}, \bibinfo
		{author} {\bibfnamefont {M.}~\bibnamefont {Tsubota}},\ and\ \bibinfo {author}
		{\bibfnamefont {J.}~\bibnamefont {Zaanen}},\ }\bibfield  {title} {\bibinfo
		{title} {Motion of a superfluid vortex according to holographic quantum
			dissipation},\ }\href {https://doi.org/10.1103/PhysRevB.107.144511}
	{\bibfield  {journal} {\bibinfo  {journal} {Phys. Rev. B}\ }\textbf {\bibinfo
			{volume} {107}},\ \bibinfo {pages} {144511} (\bibinfo {year}
		{2023})}\BibitemShut {NoStop}%
	\bibitem [{\citenamefont {Maldacena}(1999)}]{maldacena1999large}%
	\BibitemOpen
	\bibfield  {author} {\bibinfo {author} {\bibfnamefont {J.}~\bibnamefont
			{Maldacena}},\ }\href {https://doi.org/10.1023/a:1026654312961} {\bibfield
		{journal} {\bibinfo  {journal} {International Journal of Theoretical
				Physics}\ }\textbf {\bibinfo {volume} {38}},\ \bibinfo {pages} {1113}
		(\bibinfo {year} {1999})}\BibitemShut {NoStop}%
	\bibitem [{\citenamefont {Gubser}\ \emph {et~al.}(1998)\citenamefont {Gubser},
		\citenamefont {Klebanov},\ and\ \citenamefont {Polyakov}}]{gubser1998gauge}%
	\BibitemOpen
	\bibfield  {author} {\bibinfo {author} {\bibfnamefont {S.}~\bibnamefont
			{Gubser}}, \bibinfo {author} {\bibfnamefont {I.}~\bibnamefont {Klebanov}},\
		and\ \bibinfo {author} {\bibfnamefont {A.}~\bibnamefont {Polyakov}},\
	}\bibfield  {title} {\bibinfo {title} {Gauge theory correlators from
			non-critical string theory},\ }\href
	{https://doi.org/https://doi.org/10.1016/S0370-2693(98)00377-3} {\bibfield
		{journal} {\bibinfo  {journal} {Physics Letters B}\ }\textbf {\bibinfo
			{volume} {428}},\ \bibinfo {pages} {105} (\bibinfo {year}
		{1998})}\BibitemShut {NoStop}%
	\bibitem [{\citenamefont {Witten}(1998)}]{witten1998anti}%
	\BibitemOpen
	\bibfield  {author} {\bibinfo {author} {\bibfnamefont {E.}~\bibnamefont
			{Witten}},\ }\href@noop {} {\bibinfo {title} {Anti de sitter space and
			holography}} (\bibinfo {year} {1998}),\ \Eprint
	{https://arxiv.org/abs/hep-th/9802150} {arXiv:hep-th/9802150 [hep-th]}
	\BibitemShut {NoStop}%
	\bibitem [{\citenamefont {Herzog}(2009)}]{herzog2009lectures}%
	\BibitemOpen
	\bibfield  {author} {\bibinfo {author} {\bibfnamefont {C.~P.}\ \bibnamefont
			{Herzog}},\ }\bibfield  {title} {\bibinfo {title} {Lectures on holographic
			superfluidity and superconductivity},\ }\href
	{https://doi.org/10.1088/1751-8113/42/34/343001} {\bibfield  {journal}
		{\bibinfo  {journal} {Journal of Physics A: Mathematical and Theoretical}\
		}\textbf {\bibinfo {volume} {42}},\ \bibinfo {pages} {343001} (\bibinfo
		{year} {2009})}\BibitemShut {NoStop}%
	\bibitem [{\citenamefont {Hartnoll}\ \emph {et~al.}(2008)\citenamefont
		{Hartnoll}, \citenamefont {Herzog},\ and\ \citenamefont
		{Horowitz}}]{hartnoll2008building}%
	\BibitemOpen
	\bibfield  {author} {\bibinfo {author} {\bibfnamefont {S.~A.}\ \bibnamefont
			{Hartnoll}}, \bibinfo {author} {\bibfnamefont {C.~P.}\ \bibnamefont
			{Herzog}},\ and\ \bibinfo {author} {\bibfnamefont {G.~T.}\ \bibnamefont
			{Horowitz}},\ }\bibfield  {title} {\bibinfo {title} {Building a holographic
			superconductor},\ }\href {https://doi.org/10.1103/PhysRevLett.101.031601}
	{\bibfield  {journal} {\bibinfo  {journal} {Phys. Rev. Lett.}\ }\textbf
		{\bibinfo {volume} {101}},\ \bibinfo {pages} {031601} (\bibinfo {year}
		{2008})}\BibitemShut {NoStop}%
	\bibitem [{\citenamefont {Baggioli}\ and\ \citenamefont
		{Frangi}(2022)}]{baggioli2022holographic}%
	\BibitemOpen
	\bibfield  {author} {\bibinfo {author} {\bibfnamefont {M.}~\bibnamefont
			{Baggioli}}\ and\ \bibinfo {author} {\bibfnamefont {G.}~\bibnamefont
			{Frangi}},\ }\bibfield  {title} {\bibinfo {title} {Holographic supersolids},\
	}\href {https://doi.org/10.1007/JHEP06(2022)152} {\bibfield  {journal}
		{\bibinfo  {journal} {Journal of High Energy Physics}\ }\textbf {\bibinfo
			{volume} {2022}},\ \bibinfo {pages} {152} (\bibinfo {year}
		{2022})}\BibitemShut {NoStop}%
	\bibitem [{\citenamefont {Yang}\ \emph {et~al.}(2021)\citenamefont {Yang},
		\citenamefont {Xia}, \citenamefont {Zeng},\ and\ \citenamefont
		{Zhang}}]{yang2021phase}%
	\BibitemOpen
	\bibfield  {author} {\bibinfo {author} {\bibfnamefont {W.-C.}\ \bibnamefont
			{Yang}}, \bibinfo {author} {\bibfnamefont {C.-Y.}\ \bibnamefont {Xia}},
		\bibinfo {author} {\bibfnamefont {H.-B.}\ \bibnamefont {Zeng}},\ and\
		\bibinfo {author} {\bibfnamefont {H.-Q.}\ \bibnamefont {Zhang}},\ }\bibfield
	{title} {\bibinfo {title} {Phase separation and exotic vortex phases in a
			two-species holographic superfluid},\ }\href
	{https://doi.org/10.1140/epjc/s10052-021-08838-x} {\bibfield  {journal}
		{\bibinfo  {journal} {The European Physical Journal C}\ }\textbf {\bibinfo
			{volume} {81}},\ \bibinfo {pages} {21} (\bibinfo {year} {2021})}\BibitemShut
	{NoStop}%
	\bibitem [{\citenamefont {Sonner}\ \emph
		{et~al.}(2015{\natexlab{a}})\citenamefont {Sonner}, \citenamefont {del
			Campo},\ and\ \citenamefont {Zurek}}]{sonner2015universal}%
	\BibitemOpen
	\bibfield  {author} {\bibinfo {author} {\bibfnamefont {J.}~\bibnamefont
			{Sonner}}, \bibinfo {author} {\bibfnamefont {A.}~\bibnamefont {del Campo}},\
		and\ \bibinfo {author} {\bibfnamefont {W.~H.}\ \bibnamefont {Zurek}},\
	}\bibfield  {title} {\bibinfo {title} {Universal far-from-equilibrium
			dynamics of a holographic superconductor},\ }\href
	{https://doi.org/10.1038/ncomms8406} {\bibfield  {journal} {\bibinfo
			{journal} {Nature Communications}\ }\textbf {\bibinfo {volume} {6}},\
		\bibinfo {pages} {7406} (\bibinfo {year} {2015}{\natexlab{a}})}\BibitemShut
	{NoStop}%
	\bibitem [{\citenamefont {Chesler}\ \emph {et~al.}(2015)\citenamefont
		{Chesler}, \citenamefont {Garc\'{\i}a-Garc\'{\i}a},\ and\ \citenamefont
		{Liu}}]{chesler2015defect}%
	\BibitemOpen
	\bibfield  {author} {\bibinfo {author} {\bibfnamefont {P.~M.}\ \bibnamefont
			{Chesler}}, \bibinfo {author} {\bibfnamefont {A.~M.}\ \bibnamefont
			{Garc\'{\i}a-Garc\'{\i}a}},\ and\ \bibinfo {author} {\bibfnamefont
			{H.}~\bibnamefont {Liu}},\ }\bibfield  {title} {\bibinfo {title} {Defect
			formation beyond kibble-zurek mechanism and holography},\ }\href
	{https://doi.org/10.1103/PhysRevX.5.021015} {\bibfield  {journal} {\bibinfo
			{journal} {Phys. Rev. X}\ }\textbf {\bibinfo {volume} {5}},\ \bibinfo {pages}
		{021015} (\bibinfo {year} {2015})}\BibitemShut {NoStop}%
	\bibitem [{\citenamefont {Zeng}\ \emph {et~al.}(2023)\citenamefont {Zeng},
		\citenamefont {Xia},\ and\ \citenamefont {del Campo}}]{zeng2023universal}%
	\BibitemOpen
	\bibfield  {author} {\bibinfo {author} {\bibfnamefont {H.-B.}\ \bibnamefont
			{Zeng}}, \bibinfo {author} {\bibfnamefont {C.-Y.}\ \bibnamefont {Xia}},\ and\
		\bibinfo {author} {\bibfnamefont {A.}~\bibnamefont {del Campo}},\ }\bibfield
	{title} {\bibinfo {title} {Universal breakdown of kibble-zurek scaling in
			fast quenches across a phase transition},\ }\href
	{https://doi.org/10.1103/PhysRevLett.130.060402} {\bibfield  {journal}
		{\bibinfo  {journal} {Phys. Rev. Lett.}\ }\textbf {\bibinfo {volume} {130}},\
		\bibinfo {pages} {060402} (\bibinfo {year} {2023})}\BibitemShut {NoStop}%
	\bibitem [{\citenamefont {Xia}\ and\ \citenamefont
		{Zeng}(2020)}]{xia2020winding}%
	\BibitemOpen
	\bibfield  {author} {\bibinfo {author} {\bibfnamefont {C.-Y.}\ \bibnamefont
			{Xia}}\ and\ \bibinfo {author} {\bibfnamefont {H.-B.}\ \bibnamefont {Zeng}},\
	}\bibfield  {title} {\bibinfo {title} {Winding up a finite size holographic
			superconducting ring beyond kibble-zurek mechanism},\ }\href
	{https://doi.org/10.1103/PhysRevD.102.126005} {\bibfield  {journal} {\bibinfo
			{journal} {Phys. Rev. D}\ }\textbf {\bibinfo {volume} {102}},\ \bibinfo
		{pages} {126005} (\bibinfo {year} {2020})}\BibitemShut {NoStop}%
	\bibitem [{\citenamefont {Xia}\ and\ \citenamefont
		{Zeng}(2021)}]{xia2021kibble}%
	\BibitemOpen
	\bibfield  {author} {\bibinfo {author} {\bibfnamefont {C.-Y.}\ \bibnamefont
			{Xia}}\ and\ \bibinfo {author} {\bibfnamefont {H.-B.}\ \bibnamefont {Zeng}},\
	}\href@noop {} {\bibinfo {title} {Kibble zurek mechanism in rapidly quenched
			phase transition dynamics}} (\bibinfo {year} {2021}),\ \Eprint
	{https://arxiv.org/abs/2110.07969} {arXiv:2110.07969 [cond-mat.stat-mech]}
	\BibitemShut {NoStop}%
	\bibitem [{\citenamefont {Li}\ \emph {et~al.}(2013)\citenamefont {Li},
		\citenamefont {Tian},\ and\ \citenamefont {Zhang}}]{li2013periodically}%
	\BibitemOpen
	\bibfield  {author} {\bibinfo {author} {\bibfnamefont {W.-J.}\ \bibnamefont
			{Li}}, \bibinfo {author} {\bibfnamefont {Y.}~\bibnamefont {Tian}},\ and\
		\bibinfo {author} {\bibfnamefont {H.}~\bibnamefont {Zhang}},\ }\bibfield
	{title} {\bibinfo {title} {Periodically driven holographic superconductor},\
	}\href {https://doi.org/10.1007/JHEP07(2013)030} {\bibfield  {journal}
		{\bibinfo  {journal} {Journal of High Energy Physics}\ }\textbf {\bibinfo
			{volume} {2013}},\ \bibinfo {pages} {30} (\bibinfo {year}
		{2013})}\BibitemShut {NoStop}%
	\bibitem [{\citenamefont {Sonner}\ \emph
		{et~al.}(2015{\natexlab{b}})\citenamefont {Sonner}, \citenamefont {del
			Campo},\ and\ \citenamefont {Zurek}}]{Sonner_2015}%
	\BibitemOpen
	\bibfield  {author} {\bibinfo {author} {\bibfnamefont {J.}~\bibnamefont
			{Sonner}}, \bibinfo {author} {\bibfnamefont {A.}~\bibnamefont {del Campo}},\
		and\ \bibinfo {author} {\bibfnamefont {W.~H.}\ \bibnamefont {Zurek}},\
	}\bibfield  {title} {\bibinfo {title} {Universal far-from-equilibrium
			dynamics of a holographic superconductor},\ }\bibfield  {journal} {\bibinfo
		{journal} {Nature Communications}\ }\textbf {\bibinfo {volume} {6}},\ \href
	{https://doi.org/10.1038/ncomms8406} {10.1038/ncomms8406} (\bibinfo {year}
	{2015}{\natexlab{b}})\BibitemShut {NoStop}%
	\bibitem [{\citenamefont {Zeng}\ \emph {et~al.}(2021)\citenamefont {Zeng},
		\citenamefont {Xia},\ and\ \citenamefont {Zhang}}]{Zeng_2021}%
	\BibitemOpen
	\bibfield  {author} {\bibinfo {author} {\bibfnamefont {H.-B.}\ \bibnamefont
			{Zeng}}, \bibinfo {author} {\bibfnamefont {C.-Y.}\ \bibnamefont {Xia}},\ and\
		\bibinfo {author} {\bibfnamefont {H.-Q.}\ \bibnamefont {Zhang}},\ }\bibfield
	{title} {\bibinfo {title} {Topological defects as relics of spontaneous
			symmetry breaking from black hole physics},\ }\bibfield  {journal} {\bibinfo
		{journal} {Journal of High Energy Physics}\ }\textbf {\bibinfo {volume}
		{2021}},\ \href {https://doi.org/10.1007/jhep03(2021)136}
	{10.1007/jhep03(2021)136} (\bibinfo {year} {2021})\BibitemShut {NoStop}%
	\bibitem [{sup()}]{supplementary}%
	\BibitemOpen
	\href@noop {} {\bibinfo {title} {Supplementary: The movie shows the complete
			quenching process, the vortex generation process when the periodic external
			field amplitude drops from the critical value to the final quenching
			value.}}\BibitemShut {Stop}%
\end{thebibliography}

%\printbibliography{}


\end{document}